\documentclass[8pt]{article} \usepackage{amsmath} \usepackage{amssymb}
\usepackage{amsthm} \usepackage{algorithm,algorithmic}

\theoremstyle{definition} \newtheorem{defn}{Definition}[section]

\usepackage[top = 1.3in, bottom = 1.25in, left = 1.2in, right =
1.2in]{geometry}
\usepackage[T1]{fontenc} \usepackage{color} \usepackage{booktabs}
\usepackage{relsize} \usepackage[toc, page]{appendix} \usepackage{graphicx}
\usepackage{url}

\begin{document} \begin{titlepage} \centering \vspace*{2cm} {\Huge
\textbf{Privacy on the Blockchain:\\ Unique Ring Signatures.
} \par}

  \vspace{3cm} {\Huge Rebekah Mercer \par} \vspace{4cm} {\large This report is
  submitted as part requirement for the MSc in Information Security at
  University College London.~\footnote{This report is substantially the result
  of my own work except where explicitly indicated in the text.  The report may
  be freely copied and distributed provided the source is explicitly
  acknowledged.}\par}
  \vspace{2cm} {Supervised by Dr Nicolas T. Courtois.}

\end{titlepage} \newpage
\section*{\Huge\center{Abstract}}

\vspace{1cm} Ring signatures are cryptographic protocols designed to allow any
member of a group to produce a signature on behalf of the group, without
revealing the individual signer's identity.  This offers group members a level
of anonymity not attainable through generic digital signature schemes.  We call
this property `plausible deniability', or anonymity with respect to an
anonymity set.  We concentrate in particular on implementing privacy on the
blockchain, introducing a unique ring signature scheme that works with existing
blockchain systems.

We implement a unique ring signature (URS) scheme using secp256k1, creating the
first implementation compatible with blockchain libraries in this way, so as
for easy implementation as an Ethereum smart contract. 

We review the privacy and security properties offered by the scheme we have
constructed, and compare its efficiency with other commonly suggested
approaches to privacy on the blockchain.
\newpage

\section*{\Huge\center{Acknowledgements}}

\vspace{1cm} Thank you to Dr Nicolas Courtois, for introducing me to
blockchains and applied cryptography. \\

Thanks to Matthew Di Ferrante, for continued insight and expertise on Ethereum
and adversarial security. \\

Thanks also to my mum and dad, for convincing me that can handle an MSc
alongside a full time job, and for all the advice along the way! \\

Finally, thanks to the University of Manchester, UCL, Yorkshire Ladies' Trust,
and countless others for their generous grants over the years, without which
this MSc would not have been possible.

\newpage \tableofcontents \newpage

\section{Introduction} \subsection{Cryptocurrencies} Satoshi
Nakamoto\footnote{A pseudonym.} introduced the world to the proof-of-work
blockchain, through the release of the bitcoin whitepaper in 2008 \cite{bit},
allowing users and interested parties to consider for the first time a
trustless system, with which it is possible to securely transfer money to
untrusted and unknown recipients.  Since its launch, the success of bitcoin has
motivated the creation of many other cryptocurrencies, both those built upon
bitcoin's underlying structure \cite{ZeroCash}, \cite{Darkcoin}, and those
built entirely independently \cite{noether2016ring}.

Cryptocurrencies are most simply described as `blockchains' with a
corresponding token or coin, with which you can create transactions that are
then verified and stored in a block on the underlying blockchain.
\begin{figure}[h!] \centering \caption{An illustration of a typical
blockchain.} \includegraphics{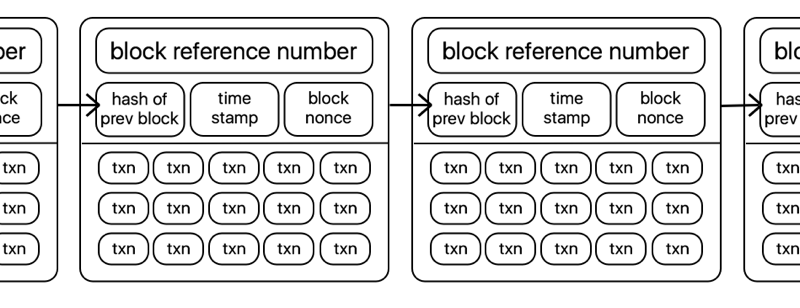} \end{figure}

Blocks themselves are formed of a block header, which includes metadata such as
the block reference number, timestamp, and a link back to the previous block;
and the block content, where the list of validated transactions is stored, each
containing the transaction value, sender and recipient address.  The blockchain
is simply the chain of blocks that have been verified and accepted as valid by
the consensus protocol in play.

Before bitcoin and the blockchain, in 1983, Chaum constructed e-cash, an
untraceable electronic alternative to cash \cite{chaum1983blind}.  Chaum's
e-cash differs from the schemes we will explore in that it assumes existence of
a central bank, from which anonymous coins can be withdrawn using \textbf{blind
signatures}.  These coins are produced with a bank's secret key, meaning that
the scheme is based on assumptions both of an honest central bank, and a
central secret \cite{chaum1983blind}.

The cryptocurrencies that we explore are intended to be analogous to digital
cash, with the same fungibility and anonymity guarantees that are implicit when
using fiat currencies such as the pound, euro, or dollar \cite{bald2015}, but
without the necessity of a central, issuing authority.

\subsection{Bitcoin}

Bitcoin is the most popular and successful cryptocurrency, with a total market
capitalisation of almost \$10 billion, as of August 2016 \cite{cmc}.

Bitcoin was formed by combining Adam Back's HashCash puzzles \cite{Back:1997}
with the established public key infrastructure, enabling the confirmation of
integrity of transactions published to the bitcoin blockchain through digital
signatures.  HashCash puzzles are used as a Proof of Work (PoW) task to be
completed by \emph{miners}, a connected network of scheme participants, in
order to make the cost of subverting or reversing transactions prohibitively
expensive \cite{bit}.

\begin{figure}[h] \centering \caption{Market Capitalisation of bitcoin (in USD)
\cite{bci}} \includegraphics[width=\textwidth]{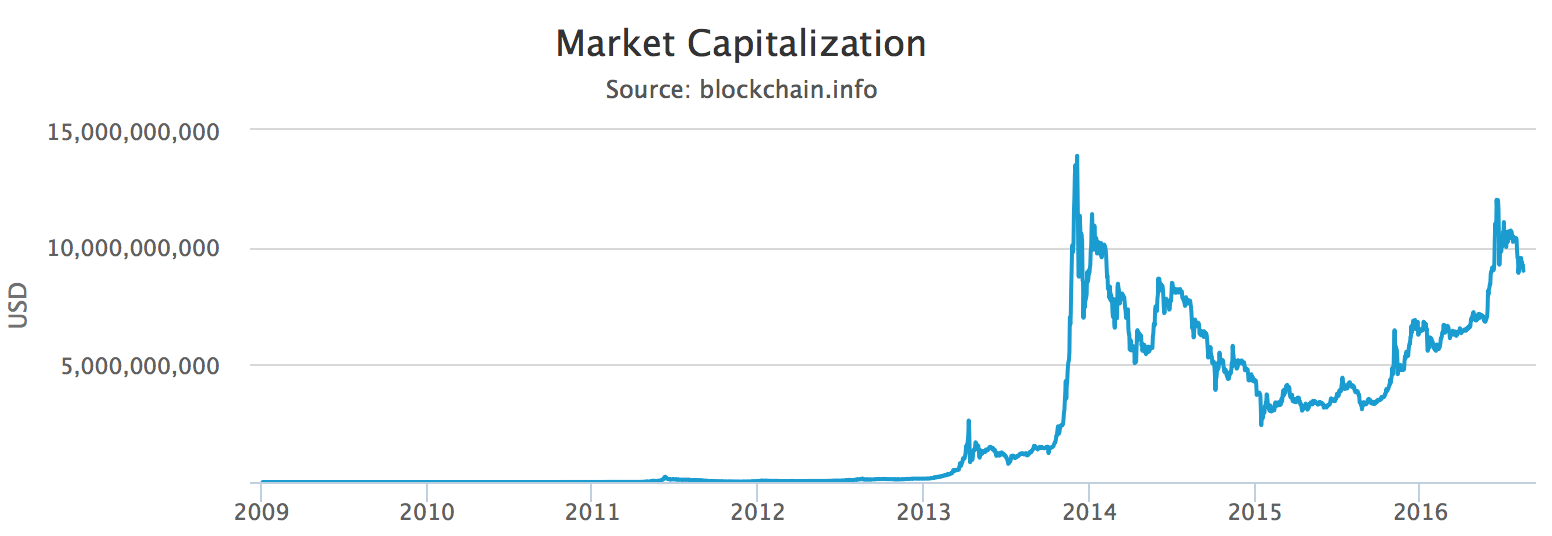} \end{figure}

Due to the nature of a PoW algorithm, in order to make a change to a previous
transaction, an adversary would require control of an amount of power equal to
the power of all those trying to contribute to the network honestly. For this
reason, it is essential that finding a solution to the given PoW algorithm is
very computationally expensive. In order for blocks to be verified quickly, it
is also generally required that given a solution, it is very efficient to
verify that it does satisfy the PoW algorithm in question.

For bitcoin\footnote{We call both the blockchain system and the unit of
currency bitcoin, with no capital letter, unless starting a sentence.}, the PoW
puzzle is formed, with || representing concatenation: \begin{equation*}
  \textbf{MSB$_k$} [\text{SHA-256}(\text{SHA-256}(\text{nonce||block
  contents}))] = 0 \end{equation*} Miners try to find a nonce such that when
  combined with the block header of the previous block, and SHA-256 (taken from
  the NIST SHA-2 suite \cite{standardnational}) is performed on the product
  twice, the output is lower than a dynamically adjusted target (above, the
  target is $2^{256-k}$) \cite{bit}.

The proof of work algorithm was an attempt to implement a `1 CPU, 1 vote'
distribution of power in the consensus \cite{bit}, and enables the system to
operate without a political or reputational requirement to be met before a
potential miner is able to contribute to the consensus algorithm -- a miner
must simply control a sufficient amount of computational power.

\begin{figure}[h] \centering \caption{The relationship between sender, miner,
and recipient. \label{txnmine}} \vspace*{2mm}
\includegraphics[width=\textwidth]{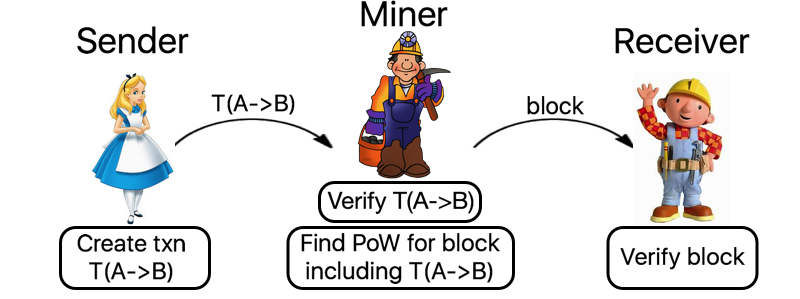} \end{figure}

The hash target is chosen in proportion to the total amount of hashing power at
work across the network, so that a new block is published approximately every
10 minutes \cite{ant2014}.  The target is adjusted every 2016 blocks
(approximately every two weeks) so that mining the next 2016 blocks should take
exactly two weeks, if the total network power does not change \cite{bitdevel}.

Miners are incentivised to perform these computationally expensive actions
through a reward of 12.5 bitcoins per block mined (this reward halves
approximately every 4 years -- the most recent halving was July 9th, 2016).  In
addition to this, transactions can include optional transaction fees to be
rewarded to the miner that includes the given transaction in a block, as
additional  motivation for miners to process certain transactions as quickly as
possible \cite{bit}.  A graph of transaction fees as a percentage of total
transaction values is shown in Figure \ref{fees} below.

\begin{figure}[h] \centering \caption{Fees as a percentage of transaction
amounts in bitcoin blocks over time \cite{bci}. \label{fees}}
\includegraphics[width=\textwidth]{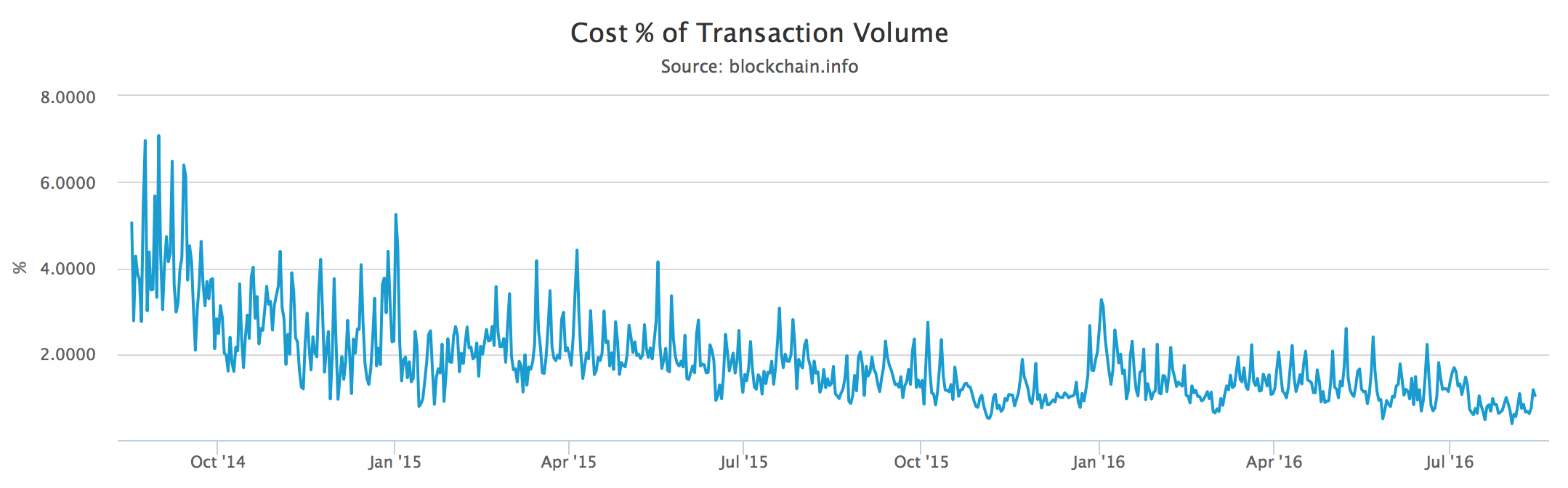} \end{figure}

\begin{figure}[h] \centering \caption{The relationship between private key,
public key, and public address. \label{rel}} \vspace*{2mm}
\includegraphics[width=0.8\textwidth]{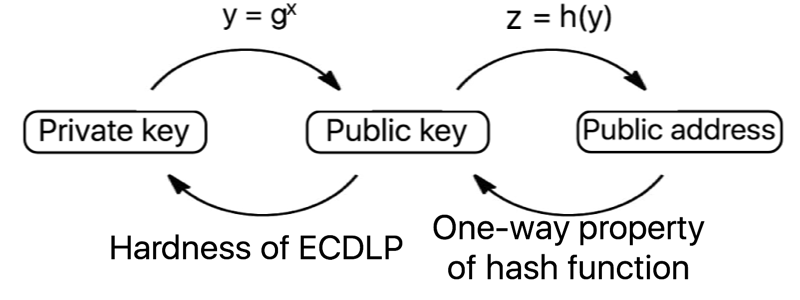} \end{figure}

Bitcoin transactions simply grant the recipient the right to spend some
currently `unspent' bitcoin belonging to the sender. This is done through
unspent transaction outputs, or UTXOs \cite{bit}.  This right is transferred
through ECDSA digital signatures, requiring use of a private key to send
bitcoin, and a public key to receive bitcoin. The recipient's public address,
formed through taking a hash of the public key, can also be used in the
creation of transactions.  The relationship between the three pieces of
information for each given user is shown in Figure \ref{rel}. We have, as
usual, private key $x$, and public key $y = g^x$, with $g$ a generator of the
group in which we are working.

 The
arrows along the bottom of the image are labelled with known computationally
hard problems.  If both of these hard problems are broken, users will be able
to calculate all pieces of information with knowledge of any one.  If ECDLP is
broken, in particular, users will be able to calculate others' private keys
from public keys. Public addresses are used as an additional layer of security
in case of this.

Users are referenced in transactions by their bitcoin addresses, of which users
can have many, generally managed by a `wallet'.  Addresses act as pseudonyms
for users.  Transactions are issued by a digital signature on the hash of
certain data concerning the current transaction, and this digital signature and
additional transaction data is then stored on the blockchain to be referenced
in future transactions \cite{Courtois:digsig}.

\subsection{Blockchains}

The blockchain can be described as an immutable, cumulative ledger, with
consensus protocol working to maintain this ledger of all valid transactions on
every node in the network.  Every transaction included in each block is public
and immutable.

There are two commonly discussed problems which a blockchain consensus protocol
must defend against: \begin{description} \item[Byzantine Fault Tolerance.]
      Consensus must hold in the presence of many faulty or badly connected
      nodes in the network \cite{Buterin:2015d}.  \item[Sybil Attacks.]
Consensus must hold even in a network with many nodes with faked or forged
identities -- generally assumed to be owned by adversarial individuals
\cite{bit}.  \end{description}

Bitcoin's protocol works towards achieving consensus under these conditions
through the slow and expensive PoW algorithm, and under the additional
requirement of an honest majority. Honest, here, means that the actor or actors
in question follow the protocol as specified.  Alternative protocols, such as a
Proof of Stake rather than PoW algorithm, are currently both implemented
\cite{NXT}, \cite{tendermint}, and being researched in industry as a way to
reach consensus more quickly and with higher security guarantees \cite{Zamfir}.

At a very high level, Proof of Stake in its current form generally works with
validators instead of PoW's miners, committing (or `staking') money in the form
of bets on certain blocks being accepted.  These stakes replace the money
miners spend on exertion of electricity in the PoW algorithm, leading to the
analogy of staking as `virtual mining'. An often cited problem with several
possible Proof of Stake algorithms is the issue of `nothing at stake', meaning
parties can act in a way where they can subvert the consensus algorithm at no
cost to themselves.

The bitcoin blockchain is equipped with a restricted programming language,
which is used to enable transactions such as transfer of funds from several
input accounts to several output accounts, including multi-signature accounts
and transactions, where several signatures are required before funds can be
released from an account.  The scripting language is expressive enough to allow
cryptographic protocols such as secure multi-party computations to be
constructed \cite{bitsecure}.

As bitcoin is primarily a transaction only blockchain, it is wise to restrict
the scripting language to a few well-examined opcodes, as this prevents risks
of unknown attacks\footnote{Without restrictions or formal verification,
scripting languages (such as Ethereum's Solidity) can lead to unexpected
attacks, such as the \$50 million stolen from the DAO, due to a formally
unknown recursive call exploit \cite{DAO}}.

There are, however, many other issues contributing to the hesitation of
widespread adoption of bitcoin and other public blockchains, including:

\begin{description} \item[The threat of attacks by an anonymous majority or
      powerful individual.] This could take place in the form of theft or
      subversion of the consensus algorithm.  Blockchain history can be
      rewritten if a dishonest party is in control of 51\% or more of the
      network \cite{bit}. Other attacks on the system can be carried out with
      control of around 25\% of the network hashing power \cite{tendermint}, or
      by groups of any size \cite{Eyal:2014kd}.  \item[Wastefulness and lack of
        sustainability.] In order to make double spending and other blockchain
        manipulation prohibitively expensive, the PoW algorithm is very
        computationally expensive, and so maintaining the blockchain consumes
        very large amounts of energy\footnote{ At \$0.125/kWh (US average),
        hash rate of 100 Mhash/s, power use of 7W, and total bitcoin hash rate
        of 1,626,365 TH/s, if all miners were using this FPGA \cite{FPGA} in
        the US` to mine, \$ would be spent on mining bitcoin every second.
        Realistically, there are FPGAs more efficient than this, and ASICs are
        10-100 times more efficient \cite{courtoisASIC}, power consumption and
        expenditure may be considerably lower.}.  \item[Low transaction rate.]
          Bitcoin's peak transaction rate is currently 7 transactions per
          second (tps) \cite{Eth}.  For comparison, VISA handles an average of
          around 2000 tps, with peak capacity of 24,000 \cite{VISA}.
        \item[Lack of control over the monetary supply, and extreme volatility
          in currency value.] Cryptocurrency value is determined by market
          forces, and uncertainty in the new technology often leads to high
          volatility. Financial institutions that would otherwise find
          blockchain technology attractive are dissuaded by the lack of control
          over the monetary creation \cite{rscoin}. This problem is resolved
          somewhat through the use of private or consortium blockchains.
\end{description}

\begin{figure} \centering \caption{Transactions per bitcoin block over time
\cite{bci}.} \includegraphics[width=\textwidth]{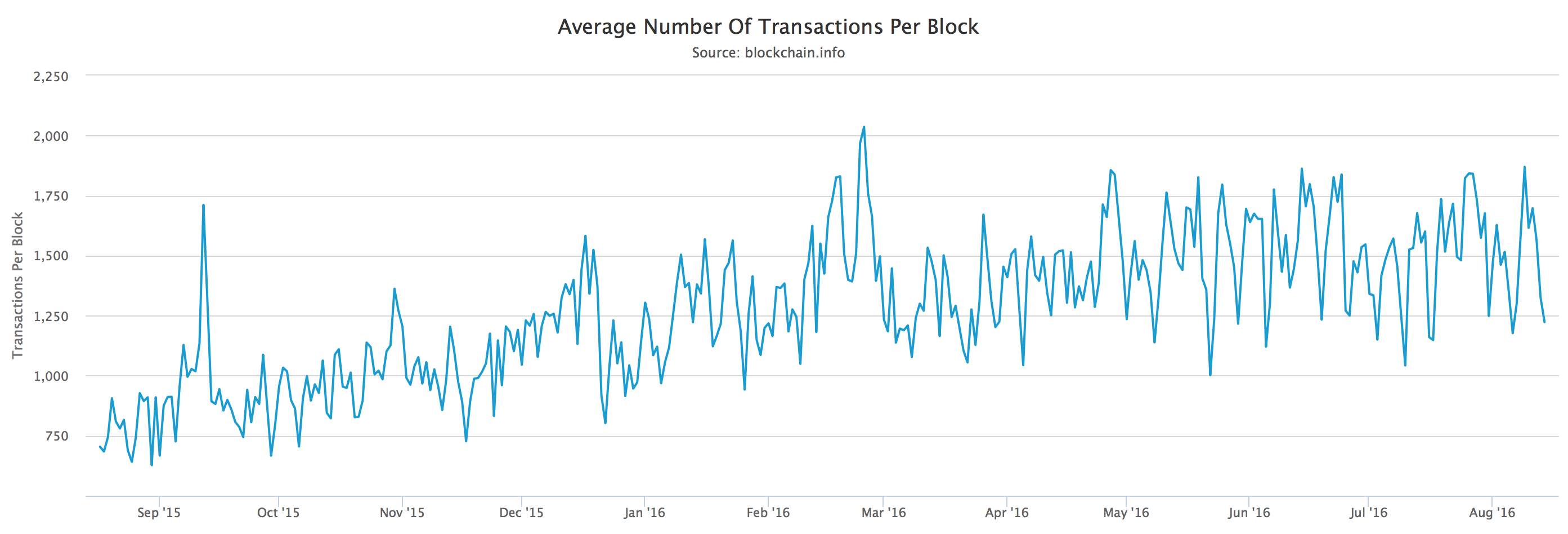} \end{figure}

\subsection{Ethereum} In response to the limitations of bitcoin's restricted
scripting language, the Ethereum platform was created, offering an
almost\footnote{Though not exactly - executions in the Ethereum virtual machine
are guaranteed to terminate, as there is an upper limit imposed on the
execution time allowed before a program must halt.} Turing-complete distributed
virtual machine atop the Ethereum blockchain \cite{Eth}, along with a currency
called Ether.  The increased scripting ability of the system enables developers
to create `smart contracts' on the blockchain, programs with rich functionality
and the ability to operate on the blockchain \textbf{state}. The blockchain
state records current ownership of money and of the local, persistent storage
offered by Ethereum.  Smart contracts are limited only by the amount of
\emph{gas} they consume. Gas is a sub-currency of the Ethereum system, existing
to impose a limit on the amount of computational time an individual contract
can use \cite{yp}.

Users specify an upper limit on the amount of gas they are willing to spend,
and the price, in Ether per gas, that they are willing to pay \cite{yp}.  An
overall block gas limit determines how many computations can be completed per
block, and although initially set as a constant, miners have the ability to
adjust the block limit by adding or subtracting the variable $\frac{1}{1024}
\times \text{previous block limit}$ to each new block's limit \cite{Eth}.

There are two types of account in the Ethereum blockchain system: `externally
owned' accounts, controlled with a private key (like all accounts in the
bitcoin system), and `contracts', controlled by the code that resides in the
specific address in question.  Contracts have immutable code stored at the
contract address, and additional storage which can be read from and written to
by the contract.

An Ethereum transaction contains the destination address, optional data, the
gas limit, the sequence number and signature authorising the transaction.  If
the destination address corresponds to a contract, the contract code is then
executed, subject to the gas limit defined above, which allows a certain number
of computational steps before halting.

The ability to form smart contracts is one potential method of addressing the
lack of privacy and corresponding potential lack of fungibility of coins in the
Ethereum system.  Although cryptocurrencies provide some privacy with the
absence of identity related checks required to buy, mine, or spend coins, the
full transaction history is public, enabling any motivated individual to track
and link users' purchases.  This concept heavily decreases the fungibility of
cryptocurrencies, allowing very revealing \textbf{taint analysis} of coins to
be performed, and leading to suggestions of blacklisting coins which were once
flagged as stolen \cite{BTCUserPriv}.

\subsection{Privacy} The blockchain privacy model is succinctly explained in
the bitcoin whitepaper, and contrasted with the traditional financial security
model as follows.  Although transactions are public, as long as public keys do
not become associated with individuals' off-chain identities, users remain
`anonymous'.  This unlinking of off-chain identities with virtual addresses and
transactions offers blockchain users a property known as \textbf{pseudonymity}
\cite{bit}.

There are certain advantages offered by a public blockchain - for example, it
is possible to track assets through the blockchain with `coloured coins'
\cite{ccs}, and certificates published on the blockchain enable collectors to
verify that they are acquiring a legitimate product.

Others have seen the public nature of transactions as a threat to fungibility
and a contradiction to the otherwise `pseudonymous' nature of bitcoin. In
reaction to this, the cryptocurrencies Darkcoin (now known as Dash
\cite{Darkcoin}), Monero \cite{noether2016ring}, and ShadowCash \cite{SDC} have
been released, and many new protocols have been suggested as privacy enhancing
overlays to the existing bitcoin system, most notably Bitcoin Core Developer
Greg Maxwell's CoinJoin \cite{coinjoin}, and the Zerocoin protocol
\cite{miers2013zerocoin}.

We explore these schemes and the cryptographic protocols they involve,
considering the practicality and security of each.

\subsection{Contributions} We produce a practical implementation of a unique
ring signature scheme\footnote{ All code related to this project can be found
at \url{https://github.com/rebekah93/secp256k1-urs/}.  }, and provide detailed
proofs and intuitive explanations of what security guarantees are provided by
the resulting scheme. We analyse cryptographic protocols and primitives that
are currently implemented in blockchain systems, and review the strengths,
vulnerabilities, and the practicality of implementing suggested improvements.
We explore hashing into elliptic curve groups, specifically implementing a
scheme to hash into secp256k1.  We work towards constructing and implementing a
scheme that enables the existence of a blockchain system that is transparent
enough to guarantee that there are no forged coins in the system, remains
auditable as a whole, but enables privacy for the individual.

\section{Background} \subsection{Cryptographic Primitives} Group signatures,
ring signatures and their variants are constructed primarily from a few well
known cryptographic primitives.  We will first describe and discuss these
primitives, so we can more easily describe the protocols and requirements of
the schemes we deal with further on in this paper.

\subsubsection{Elliptic Curve Cryptography}

Until the introduction of Elliptic Curve (EC) cryptography and the Elliptic
Curve Discrete Logarithm Problem (ECDLP) by Koblitz and Miller in 1985,
cryptographic protocols were defined over multiplicative groups of finite
fields.

The advantages of using EC groups rather than finite fields include the
efficiency and speed of EC arithmetic, and the absence of any sub-exponential
time algorithms with which to find the discrete logarithm in an EC group.  The
ECDLP in the group $E ( \mathbb{F}_q )$ is believed to be strictly more
difficult than the DLP in finite fields of size $q$ \cite{petit2016algebraic}.

An elliptic curve, labelled $E(\mathbb{F}_q)$, (with
characteristic\footnote{Characteristic is defined as the number if times the
identity element must be added to itself to return the zero element.} $\neq$ 2,
3) consists of all the points $(x,y) \in \mathbb{F}_q \times \mathbb{F}_q$ that
satisfy the following equation: \begin{equation} y^2 = x^3 + ax + b, \ \qquad \
\ \ \ \qquad \ \ \ \ a, b \in \mathbb{F}_q.  \end{equation}

$\mathbb{F}_q$ is known as the \textbf{base field} of the group, and the
cardinality, or order, of the EC group, represented $n = \#E(\mathbb{F}_q)$, is
defined as the total number of points $(x, y) \in E(\mathbb{F}_q)$.

The order of the curve also is the smallest number such that for a generator $g
\in E(\mathbb{F}_q)$,  we have $g^n = 1$.

Arithmetic in elliptic curve groups does not occur as one may naively assume,
with $(x_1, y_1) + (x_2, y_2) = (x_1 + x_2, y_1 + y_2)$. Instead, point
addition takes place through the formulae given below -- the equations used to
perform point doubling and scalar multiplication follow from the point addition
formula, and are given in Appendix B. A full description can be found in
\cite{menezes2012elliptic}

\begin{description} \item[Point Addition] For simple point addition, $(x_1,
      y_1) + (x_2, y_2) = (x_3, y_3)$, with $(x_1 \neq x_2)$, we form $(x_3,
      y_3)$ through the following: \begin{align} \text{For } \lambda & =
        \frac{y_2 - y_1}{x_2 - x_1}, \\ x_3 & = \lambda^2 - x_1 - x_2, \\ y_3 &
      = \lambda(x_1 - x_3) - y_1.  \end{align} It is clear here that $\lambda$
      is formed by taking the gradient of the line joining the two points.
      Point doubling replaces $\lambda$ with the tangential gradient at the
      point, as explained in Appendix B.  Scalar multiplication occurs through
      repeated point doubling.

It is useful to note that is EC groups are mathematical groups, they have only
    one operation -- EC groups are additive groups, meaning although
    multiplication by a scalar can be computed through repeated point
    doublings, there is no multiplication of two points, no exponentiation and
    no division. As the schemes we implement are using multiplicative notation,
    it is important to be aware of this. Multiplication in suggested algorithms
    will become point addition, and exponentiation will become scalar
    multiplication.

\item[ECDSA] Transactions in blockchain systems such as bitcoin and Ethereum
  use the Elliptic Curve Digital Signature Algorithm (ECDSA) as authentication
    of a transaction \cite{bit}.  ECDSA signatures are formed of pairs, $(r,
    s)$, constructed as shown in Appendix B, and explained in full detail in
    \cite{johnson2001elliptic}. At a high level, these digital signatures fit
    into the scheme as shown in Figure \ref{txnmine}, on page 8.

\end{description} \subsubsection{Assumed Hard Elliptic Curve Problems} As we
have seen above, elliptic curve points are additive groups, in contrast to
finite fields which are generally defined over multiplicative groups.  This
means the notation used when implementing and discussing schemes over elliptic
curve groups differs from the notation given in the papers suggesting the
algorithms.  We will informally define the most commonly used cryptographic
assumptions using the additive notation of elliptic curve arithmetic, and then
formally explore the assumption that the scheme we use depends on.  All
definitions are defined with respect to an adversary $\mathcal{A}$.

\begin{defn}\textbf{Elliptic Curve Discrete Logarithm Problem (ECDLP)}

  $\mathcal{A}$ has no advantage in solving following:

  Given $Y$, $G \in E(\mathbb{F}_q)$, find $x \in \mathbb{Z}$ such that $Y = x
\cdot G$.  \end{defn}

\begin{defn}\textbf{Computational Diffie-Hellman Assumption (EC-CDH)}

  $\mathcal{A}$ has no advantage in the following:

  Given $a \cdot G$, $b \cdot G, \in E(\mathbb{F}_q)$, $a, b \in \mathbb{Z}$,
compute $ab \cdot G$.  \end{defn}

\begin{defn}\textbf{Decisional Diffie-Hellman Assumption (EC-DDH)}

  $\mathcal{A}$ has no advantage in the following:

  Given $a \cdot G$, $b \cdot G$, $c \cdot G \in E(\mathbb{F}_q)$, with $a$,
$b$, $c \in \mathbb{Z}$, decide whether $c \cdot G = ab \cdot G$.  \end{defn}

\subsubsection{Zero-Knowledge Proofs} Goldwasser et al.\ introduced
`zero-knowledge proof of knowledge' schemes in 1985 \cite{Gold1985}.  The
purpose of the Zero-Knowledge Proof (ZKP) of knowledge is for a party to prove
to a verifier that they know some secret information (represented below as $x$)
without revealing anything about the secret in the process.

\subparagraph{Goldwasser's Scheme:} Prover and verifier both know $(g, h, y_1,
y_2)$, with $g, h \neq 1$, $g, h \in \mathbb{G}$, $y_1 = g^x$, $y_2 = h^x$, for
exponent $x \in \mathbb{Z}_q$.  The scheme runs as follows \cite{Gold1985}:
\begin{enumerate} \item Prover chooses $r \leftarrow_R \mathbb{Z}_q$, sends $a
    \leftarrow g^r$, $b \leftarrow h^r$ to the verifier.  \item Verifier
    responds with challenge $c \leftarrow_R \mathbb{Z}_q$ for the prover.
  \item Prover responds, sending $t \leftarrow r - cx$ mod $q$ to the verifier.
\item Verifier accepts if and only if $a = g^t y_1^c$ and $b = h^t y_2^c$
\end{enumerate}
For $g$ and $h$ in a group $\mathbb{G}$ where the discrete logarithm problem is
assumed hard, the scheme above is both sound and honest-verifier
zero-knowledge.

A \textbf{zero knowledge} property would guarantee that no malicious verifier
can extract additional information from the prover. \textbf{Honest verifier
zero knowledge}, a weaker property, guarantees that if the verifier follows the
protocol honestly, and chooses the challenge randomly, the zero knowledge
property follows.

Proof of knowledge, as achieved in this scheme, is a property stronger than
soundness. Proof of knowledge guarantees not only that if the verifier is
convinced a \textbf{witness} exists (above, $r$ is the witness), but also that
the prover knows such a witness \cite{ben2015secure}.  The property of
soundness, in contrast, simply guarantees that it is impossible to prove a
false statement.

Unique ring signatures rely heavily on Non-Interactive Zero Knowledge (NIZK)
proofs. In contrast to the original zero-knowledge proofs, NIZK proofs require
only that the two parties have access to a randomly generated `common reference
string'.

The Fiat-Shamir transform converts a three round ZKP with interaction and
public randomness needed to determine the challenge into a one round,
non-interactive zero-knowledge proof of knowledge, with a hash function
modelled as a random oracle.

\subsubsection{Random Oracle Model} A random oracle (RO) is generally viewed as
a `black box' accepting inputs and generating truly random outputs for them.
The random oracle records the outputs corresponding to each input queried, so
that the same input will always return the same output.  The \textbf{random
oracle model} is an assumption that there exists a random oracle producing a
truly random output.

Random oracles are typically modelled with hash functions in practice, leading
to some criticism -- hash functions have a deterministic output and so are not
truly random, security proofs in the ROM may not translate to security in a
practical environment \cite{groth2007fully}.

The security properties of all linkable, traceable, and unique ring signatures
are proved in the random oracle model, including the one we have
implemented\footnote{There are, however, some group and ring signatures that
exist without the random oracle model \cite{jens}}.

The ROM is at the foundation of the security of our scheme, and many others. As
an example, the security of bitcoin mining is dependent on hash functions
acting as random oracles -- unpredictable, well distributed, (pseudo) random
functions.  Similarly, NIZKs are dependent on the common reference string being
modelled as a RO -- otherwise a dishonest prover would have a non-negligible
chance of correctly predicting the challenge and forge a proof without having
knowledge of the secret.

\subsection{Group Signatures}
Group signatures were introduced by Chaum and van Heyst in 1991 \cite{GSig}.
They were conceived to allow any member of a group to produce a signature on
behalf of the group, enabling users to sign with the authority of the group,
without revealing the specific signer's identity.

The concept of a group signature is that a trusted group master or manager is
responsible for setting up a `group' of users, (not related to the abstract
mathematical structure of groups -- we simply mean a collection of users of the
scheme), who can then each sign messages on behalf of the whole group, without
revealing their individual identity.  The group master holds a master key with
the ability to reveal the signer of any signature generated by a group member
in the past.  As a result of this, group signatures offer the participants
anonymity only under the condition that the group master does not choose to
reveal the signer's identity  \cite{camenisch1999}.

A group signature scheme must satisfy several essential properties:
\begin{description} \item[Anonymity] An adversary has no more than a negligible
    advantage of correctly identifying the individual that produced the
  signature.  \item[Unforgeabilty] An adversary without a key has no more than
    a negligible probability of producing a signature that verifies correctly.
  \item[Collusion resistance] Dishonest participants in the group cannot
    collude to produce a signature which will verify as another's signature,
    and the scheme must offer soundness and correctness under the signature
    verification algorithm.  \end{description}

Consider the definition of anonymity with respect to a two-party group. Less
than a negligible advantage would mean here that an adversary would guess
correctly which individual produced the signature with probability $\frac{1}{2}
+ \varepsilon$, for some negligible $\varepsilon$. This means that although the
adversary has negligible advantage, the \textbf{anonymity set} is small and so
the definition of anonymity does not align with our intuitive sense of the
word.  We therefore call this property plausible deniability. Each signer can
deny that they produced the signature, but the property of `anonymity' that is
offered does not agree with ones intuitive definition of the word.

Although we will avoid detailing a group signature scheme (thorough
explanations can be found in \cite{GSig}, \cite{camenisch1999},
\cite{delerablee2006dynamic}), we will describe possible uses of group
signature schemes.  For example, a bank could allow its employees to authorise
transactions by signing via a group signature protocol.  This offers the
employees privacy from eavesdropping third parties, and reveals nothing to the
recipient other than that it was a bank employee that authorised the
transaction.  However, for auditability or in case of a dispute, the group
master can reveal the identity of the signer in each disputed or audited
transaction.

Another example is of Direct Anonymous Attestation (DAA), which uses a variant
of group signatures, to enable a server to authenticate a trusted platform
running remotely on an authorised user's laptop, without compromising the
individual user's privacy by requiring their identity \cite{DAA}.

However, group signatures have several limitations.  For example, it is
impossible for third parties to know whether or not a series of messages have
been signed by the same person -- the only way to link messages would be
through the group manager, who would have to revoke the anonymity of the signer
in the process.  The linking of messages is an essential property in any mixing
scheme, as it can be used to provide confirmation of whether or not a group
signer has already withdrawn the coins that they have rightful access to.
There is also a large overhead when attempting to add or remove members from an
established group, with the group manager or all group members required to
perform a computationally expensive task.  The presence of a trusted group
manager also restricts possible use of the protocol.

\subsection{Ring Signatures} Ring signatures were first suggested by Rivest et
al, who introduced the RST scheme in their 2001 paper, `How to Leak a Secret'
\cite{2001leak}.  Ring signatures were created in response to the limitations
of group signatures, and in particular they offer honestly participating users
with `unconditional anonymity', and are formed without a complex setup
procedure or the requirement for a group manager.  They simply require users to
be part of an existing public key infrastructure \cite{SLRSR}.

Ring Signature Mixing Schemes (RSMSs) allow different sets of blockchain users
to generate groups and signatures on the fly, without requiring any additional
trust, at the cost of little added computational time.

Ring signatures are constructed in a way that the ring can only be `completed',
and therefore verify correctly, if the signer has knowledge of some secret
information, most commonly a private key corresponding to one of the public
keys in the ring.  In the signature generation algorithm, a number is generated
at random for each of the other public keys in the ring, and then the signer
uses the knowledge of their own private key, or some other `trapdoor
information', to close the ring.

Ring signatures offer users a type of anonymity by hiding transactions within a
set of others' transactions.  If there are many users contributing very similar
amounts to the ring, then the ring is said to have good \emph{liquidity},
meaning the transactions can occur quickly, and also that transactions can be
effectively mixed, with a high resistance to attempted mixing analysis attacks.

\subsection{Linkable Ring Signatures} Linkable ring signature algorithms
provide a scheme that allows users to sign on behalf of a group, again without
revealing the individual signer's identity, but with the additional property
that any signatures produced by the same signer, whether signing the same
message or different messages, have an identifier, called a tag, linking the
signatures.  With this tag, third parties can efficiently verify that the
signatures were produced by the same signer, without learning who that signer
is.

\subsubsection{Unique Ring Signatures} Unique ring signatures have a tag that
links signatures if and only if the signer, message, and ring are the same
across the two signatures.  This tag is constructed using the signer's private
key, message, and description of the ring (most commonly a list of public
keys), and enables both other ring members and third parties to observe whether
or not two identical messages have been signed by the same ring member.

Possible use cases for linkable ring signatures are restricted access archives,
for example a journalist may pay for access to one query from each of a range
of topics. By creating a unique ring signature with the topic as message, this
would enable the archive server to allow the journalist appropriate access,
without compromising the individual's privacy.  Extending this example, we
would create `$k$-times anonymous access' \cite{teranishi2004k}. Other
potential uses for unique ring signatures including e-voting schemes without
the need for central authorities, and other e-token systems \cite{f2012}.

Unique ring signatures were introduced by Franklin and Zhang in 2012
\cite{f2012}.  The security properties of this scheme are stated as
\textbf{unforgeability, secure linkability, and restricted anonymity} -- a
fully anonymous linkable ring signature scheme would be impossible, due to the
linkability itself.  These security properties are guaranteed by the collision
resistance of the hash functions involved, and the soundness and completeness
of the zero-knowledge proof.

The unique ring signature (URS) scheme results in the ``most efficient linkable
ring signature in the random oracle model, for a given level of provable
security'' \cite{f2012}.  Specifically, this scheme is tightly reduced to the
DDH problem\footnote{Tight in this setting means that the probability of
solving the hard problem on which the security of the scheme relies is roughly
equal to the probability of solving the DDH problem in the same given time
period.}.

Franklin and Zhang build the security proof of their suggested scheme upon
security proofs of other, well-established schemes, starting from a generic
digital signature scheme, and ending with an abstract unique ring signature
over bilinear pairings \cite{f2012}.

\section{Motivation}

The motivation for increased privacy in cryptocurrencies is self-evident.  We
wish to enable cryptocurrencies to be used with a guarantee of the same level
of anonymity that users take for granted with cash -- it can be withdrawn from
a bank without the user needing to reveal their intent, and can be spent
without the merchant learning the payer's identity.  This is not yet offered by
any cryptocurrency or e-cash scheme, and certainly is not offered by credit
card purchases.

The most commonly considered options for implementing privacy on the blockchain
include the following: \begin{description} \item[Homomorphic Encryption (HE)]
      All transactions are encrypted in a way that allows transactions to be
      easily audited, without revealing the actual value of such transactions.
      Fully HE is currently very inefficient, with a billion factor overhead
      and keys up to 25GB in size \cite{gentry2009fully}.

    Confidential Transactions (CT), used to hide the value of a given
    transaction, are implemented using additively homomorphic encryption, which
    has less of a computational overhead \cite{noether2016ring}.

  \item[Mixing services] Coins can be mixed using third party servers.  An
    example mixing scheme is shown in Figure \ref{mixmix}.  The CoinJoin scheme
    has been implemented many times, as it is one of the only mixing solutions
    available to use without a modification to the bitcoin protocol.  An
    example mixing service, based on the CoinJoin protocol, is SharedCoin,
    which was hosted by blockchain.info until June 2014 \cite{sharedcoinbreak},
    when it was found to be broken \cite{sharedcoinbreak2}.  Alternatively, all
    participating parties can execute part of the mixing contract, eliminating
    the need for trusted servers.  An example of this is JoinMarket, where
    requests for mixing orders are broadcast over Internet Relay Chat (IRC)
    \cite{jm}.  In July 2015, it was discovered that JoinMarket was incorrectly
    implemented, enabling curious parties to perform analysis to link input and
    output address pairs \cite{jmb}.

    \begin{figure}[h!] \centering \caption{A simple mixing contract, used to
    improve the fungibility of coins.\label{mixmix}} \includegraphics{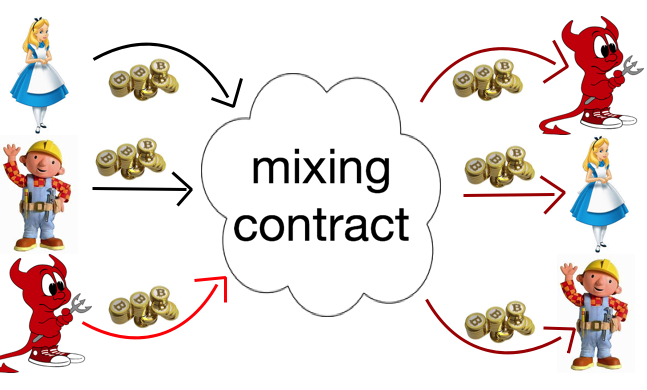}
    \end{figure}

  \item[Cryptocurrencies with privacy by design] This could include, for
  example, an obligation to enter a mix when performing any transaction -- the
cryptocurrency may not recognise transactions with only one input. Such a
cryptocurrency could also produce all transactions in `zero-knowledge', such as
ZCash \cite{hopwood}.  \item[Access structures] We can increase privacy on the
  blockchain by concealing metadata, or having transactions only readable to
    certain parties determined through access control mechanisms.
  \item[Traitor tracing] This could be implemented in order to punish parties
    who act to decrease anonymity of others, in turn increasing the anonymity
    of the scheme as parties are less likely to act adversarially.
  \item[Broadcast encryption] This provides a high level of anonymity for
    receivers, as every user in the group receives the encrypted message,
    although only users with the correct permission or key can decrypt.
    Broadcast encryption is also perfectly collusion resistant
    \cite{boneh2005collusion}.  \item[Secure hardware] This can act to limit
      what the user can do, for example the hardware could enforce mixing
      before outputting transactions, or could add randomness such as that used
      in blind signatures \cite{chaum1983blind}.  \item[Intermediaries] For
      example, depositing and withdrawing money from an exchange removes the
    `traceability' property.  \item[Secure Multi-Party Computation (SMPC)] This
      enables parties to act together in a way that no single one of them has
      access to all of the data, and hence no one can leak any secret
      information. However, inefficiencies are significant and parties are
      required to behave honestly \cite{enig}.  \item[Off-chain storage] This
        would increase privacy on the blockchain by storing sensitive data
        off-chain and simply accessing it when needed \cite{enig}.  Either the
        storage hosts are trusted, or data is fragmented and stored across many
        nodes.  \end{description}

\subsection{What could go wrong?}

Mixes provide only \textbf{plausible deniability} -- the transaction, sender
and recipient addresses are all still public, but are no longer obviously
linked, and rather exist as if `hidden in a crowd'. The size of the crowd, or
anonymity set, depends on parameters chosen either by the mixing scheme itself
\cite{jm}, or by the user selecting an anonymity level they are comfortable
with.

Due to the definition of anonymity, ring mixes with group size 2 have been
proven to provide the user with `anonymity', as long as the underlying ring
signature protocol is provably secure \cite{BKM}. However, in a blockchain
system, this may not be an adequate level of privacy -- additional information
is often available and could be used to construct a persona or perform other
revealing analysis on each member of the ring, enabling a motivated adversary
to trace the signers as with the pseudonymity that bitcoin and Ethereum
provide.

It is important to note that mixed based protocols rely on a large number of
honestly participating users, in order to offer a desirable level of anonymity.
We define anonymity here with respect to an \textbf{anonymity set}, with a
higher number being more desirable and offering `stronger' properties of
anonymity.

In the two-party ring, the anonymity set is of size two. If a user wanted a
higher level of anonymity, they could enter into a chain of mixes, each of
which with a size two, to reach the preferred level of anonymity.  For example,
entering into eight two-party ring mixing schemes would give the user an
anonymity set of $2^8$, as an adversary would have to deduce which of the two
parties is the individual in question, for each of the eight rings.

Ring signatures are not produced instantaneously, and so it is preferable to
allow users to enter into one large ring, rather than a chain of smaller rings.

\subsubsection{The Threat of Sybil Attacks}

It is not unreasonable to suggest a motivated attacker may chose to create many
accounts and flood many mixing contracts with the intent to publish all of
their public and private keys pairs after the legitimate party has withdrawn
their funds, hence revealing the honest input output address pair. This attack
would be very straight-forward to execute, and a simple overview is shown in
the figure below.

\begin{figure}[h] \centering \caption{An adversary acting as all but one users
in a mix} \label{fig:syb} \includegraphics[width=\textwidth]{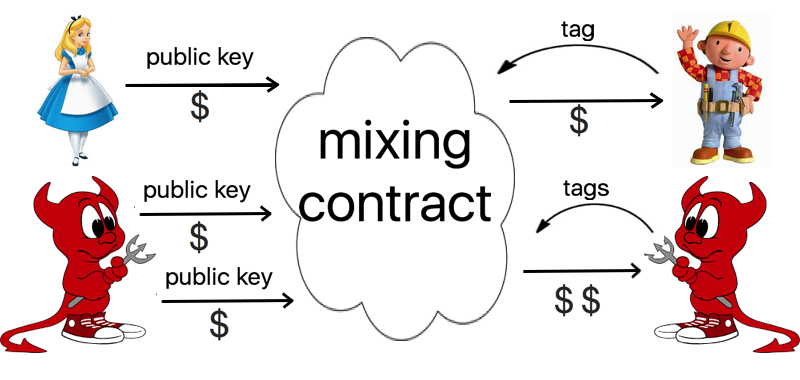}
\end{figure}

Although ring signatures offer an increased level of anonymity to users,
dishonest participants can remove themselves from the set of possible signers
of a formerly signed message by simply revealing their own private key or tag
based on the specific message and ring in question.  This would allow third
parties to trivially observe that the given user is not the signer of the
original message.

 If all but one users in a ring choose to act adversarially and reveal their
 private key or tag, then the remaining user is left only with the
 `pseudonymity' they would receive using bitcoin or a similar cryptocurrency.

 For a ring signature scheme, a straight-forward way to prevent an adversarial
 ring from revoking the exculpability of any honest member would be to
 implement a scheme with blinded signatures.  Chaum's blinded signatures, first
 suggested in 1983 \cite{chaum1983blind}, work as follows.

 For a signer with private key $x$ and public key $Y = x \cdot G$, the message
 is signed with key $Y + P$, with $P = p \cdot G$, for some randomly generated
 $p \in \mathbb{Z}_q$.  The signer also sends a zero knowledge proof of
 knowledge of $p$.

 To produce the ring signature, the signer produces a random $r \leftarrow_R
 \mathbb{Z}_q$ for each public key in the ring, and uses these to form $pk_i +
 R_i$, with $R_i = r_i \cdot G$ as the new public keys to be used to form the
 ring.

 Zero knowledge proofs of each of these random blinding constants are also
 produced and sent with the signature to the verifier.

 If this is done correctly, the non-signing parties will not be able to
 generate tags with which to compare and prove that they are not the signer in
 question, as they will not be aware of the $R_i$ used in conjunction with
 their pubic key.  If the signer also forgets the blinding constants, they will
 not be able to incriminate themselves even if asked to. However, it is unclear
 whether this scheme would work with linkable or unique ring signatures.

 \subsubsection{Failed Mixing Attempts}

Dash and SharedCoin both relied on third parties to host implementations of the
CoinJoin mixing protocol.  For Dash, then known as DarkCoin, nodes were not
willing to host the `DarkSend' mixing feature, and so five `masternodes' were
chosen to provide mixing services to the whole network, and in turn these nodes
were promised a reward.  These five nodes form an easy and attractive target
for attacks, and could have chosen to collude and act adversarially, logging or
publishing all the information collected about the input-output pairs in each
mix.  The service is now called PrivateSend, and all nodes on the network are
required to run the mixing protocol. In exchange for this service, 45\% of all
block rewards go to the nodes that provided the mixing.

SharedCoin was a service hosted by blockchain.info, which has also been broken.
In this case, an eavesdropping adversary could easily trace input-output pairs,
based on transaction values and analysis explained in CoinJoin Sudoku
\cite{sharedcoinbreak2}.

In JoinMarket \cite{jm}, all users wishing to join the mix broadcast their
intentions over an IRC channel, and then all act as the mixing server. There
is, however, a security problem with these mixing schemes, under the threat
model of an active adversary who sets up rings and simply collects messages and
input-output address pairs, without ever executing the mixing protocol. This
adversary can collect valuable information about transactions while never
executing the mixer, meaning the information is collected at no cost
\cite{jmb}.

\subsubsection{Low Power Problems} It is easy to see that the more legitimate
hashing power contributing to a given network, the more computationally
expensive an attack would be.  If this expense is large enough, an attacker
would be crippled trying to gain a sufficient\footnote{Formerly it was thought
that an attack required control of 51\% of the network
\cite{bradbury2013problem}, more recently it has been shown that adversaries
can negatively influence the system with 34\% or less \cite{Eyal:2014kd}}
amount of hashing power, and so the honest majority will cause the blockchain
to grow legitimately.

However, for some smaller alt-coins, the expenditure required to gain a
sufficient amount of hashing power is not large enough to prohibit a motivated
attacker from taking over the network.  This was a high risk in 2014 for
Litecoin and Dogecoin, which use the same hashing algorithm.  If either
system's miners had decided to collude and attack the other blockchain, they
could do so with devastating affects \cite{Courtois:2014uw}.

On July 20th, 2016, the Ethereum blockchain successfully completed a hard
fork\footnote{A change to the accepted blockchain protocol -- in this case, it
was an irregular state change in order to return stolen funds to their original
account} \cite{DAO}.  ETC (Ethereum Classic -- the chain that did not fork) has
around 16\% of the hashing power of Ethereum, which leads to a vulnerability in
the ETC system. If the Ethereum miners wanted to attack the ETC network, they
could do so very easily.

The security that comes with the network effect and higher hashing power of
existing cryptocurrencies is the main reason why creating our own private
cryptocurrency is not the best choice -- we choose instead to leverage the
power of existing blockchain platforms.

\subsection{Anonymous Alt-Coins} \subsubsection{ZCash and zk-SNARKs}
Zero-Knowledge Succinct Non-Interactive ARguments of Knowledge (zk-SNARKs) give
cryptocurrency users the ability to hide all transaction data
\cite{ben2013snarks}.  Zk-SNARKs require the sender to produce a proof, in
zero-knowledge, of the ability to spend an amount greater than or equal to the
value of the transaction they are submitting.  Zk-SNARKs satisfy perfect
completeness and computational soundness properties, in addition to the
property of succinctness, which simply means that the proof is polynomial in
the security parameter.

The proofs are currently computationally expensive to produce
\cite{cryptoeprint:2014:595}, taking several minutes to generate a key pair,
and several more to generate a zero knowledge proof, but this is an active
research area.  The proofs are, however, very computationally efficient to
verify, and the signatures are very small in size -- 322 bytes for the
signature, and ``a few milliseconds'' for the verification \cite{CSNARK}.

ZCash is a cryptocurrency which attempts to address the lack of privacy in
blockchain transaction through use of zk-SNARKs.  ZCash uses zk-SNARKs over a
merkle tree that contains all previous transactions.  Used naively, this could
leak information about the latest transaction, as the transaction would be in
the most recent set but not the previous one. To prevent this, the merkle tree
is of constant size, $2^{64}$ \cite{hopwood} -- for scale, if ZCash
transactions occur at an average rate of 1000 transactions per second, it would
take around 292 million years for the merkle tree to be filled.

ZCash is based on Zerocash which itself is an improvement of a protocol called
Zerocoin, designed by a team of researchers at John Hopkins University in 2013
\cite{miers2013zerocoin}.  Zerocoin was designed to work atop the bitcoin
blockchain system, with transaction origin hidden through a process involving
\textbf{minting} and \textbf{pouring} of coins off the bitcoin blockchain and
into zerocoins to obfuscate the otherwise transparent bitcoin transactions.

Zerocoin was implemented as an extension to bitcoin, but was not commonly used,
due to efficiency problems and imperfections of the scheme \cite{ZeroCash}.
Most importantly, using Zerocoin requires a 25kb proof to be produced for every
coin spent, resulting in a total transaction size of 49kB. The transactions
also only exist in single-denomination values, meaning many large proofs must
be computed, and multiple transactions sent, in order to transfer a large
number of zerocoins.

In 2014, the scheme was extended and renamed Zerocash, a full cryptocurrency
that used zk-SNARKs to protect the transaction value, sender and recipient
address \cite{ZeroCash}.

ZCash has since been developed as an independent cryptocurrency with a further
refined version of the Zerocash protocol \cite{hopwood}.  The properties that
each scheme offers, in addition to the properties offered by ``Zoe'', or ZCash
on Ethereum, are summarised in Table \ref{zstuff} below\footnote{The Zerocoin
operation time does not include the time required to compute the accumulator
used, explained in \cite{miers2013zerocoin}.}.

\begin{table}[h] \centering \caption{A comparison of the zk-SNARK based
  protocols.} \label{zstuff} \begin{tabular}{c c c c c}\toprule & Zerocoin
    \cite{miers2013zerocoin} & Zerocash \cite{ZeroCash} & ZCash \cite{hopwood}
    & ZoE \\ \midrule Year Proposed & 2013 & 2014 & 2016 & 2016 \\ Hides origin
    & $\checkmark$ & $\checkmark$ & $\checkmark$ & $\checkmark$ \\ Hides
    recipient & $\times$ & $\checkmark$ & $\checkmark$ & $\checkmark$ \\ Hides
    value & $\times$ & $\times $ & $\checkmark$ & $\checkmark$ \\ Blockchain &
  Bitcoin\footnote{With substantial modifications.} & Zerocash & ZCash &
  Ethereum \\ \bottomrule \end{tabular} \end{table}

The property of succinctness allows these schemes to produce proofs that are
  `small' and quickly verified in comparison with other zero-knowledge proofs
  \cite{ZeroCash}.  However, when compared with other blockchain transactions,
  the proofs and resulting transactions are very large, and there are stages
  which take a considerable amount of time.  In Table \ref{zeff}, we compare
  performance and cost of the zk-SNARK based cryptocurrencies with Ethereum and
  the original performance of SNARKs.  \begin{table}[h] \centering \caption{A
    comparison of efficiency of zk-SNARK based protocols.} \label{zeff}
    \begin{tabular}{c c c c c c}\toprule & SNARKs for C \cite{ben2013snarks} &
      Zerocoin \cite{miers2013zerocoin} & Zerocash \cite{ZeroCash} 
      & Ethereum \\ \midrule Txn size (bytes) & 322 & $\sim$49,000 & 996 & 65
      \\ Security level (bit) & N/A & 80/128 & 128 & 128 \\ Key gen time (mins)
      & 20 & ? & 5 & 0 \\ Prove time (mins) & 22 & `a few' & 1 & 0 \\ Verify
      time & 4.68 secs & 450 ms & 5.4 ms & ? \\ \bottomrule \end{tabular}
  \end{table}

All of the schemes based on zk-SNARKs at present require a \textbf{trusted
  setup stage} -- without this, the secret information used in the construction
  of the system can be used dishonestly to fake transactions and create new,
  counterfeit coins.  However, even with knowledge of the secret information,
  user anonymity cannot be compromised under any threat model, due to the
  zero-knowledge property of zk-SNARKs.  A multi-party setup is being
  considered as a way to mitigate against the necessity of a trusted setup
  stage \cite{ben2015secure}. Using N parties, rather than one trusted party,
  would offer the scheme the stronger property of security under just one-of-N
  of the involved parties acting honestly.

\subsubsection{Monero and CryptoNote}

Monero describes itself as an anonymous cryptocurrency, and uses the CryptoNote
  protocol to implement mixing through ``Multi-layered Linkable Spontaneous
  Anonymous Group Signatures (M-LSAGS) \cite{CryptN}.

As of August 2016, Monero has a total market cap of \$57,000,000, and a daily
  transaction volume of \$8,000,000, almost double that of Ethereum, at
  \$4,500,000 \cite{cmc}, both of which are only a fraction of bitcoin's
  \$58,000,000.

At present, Monero has implemented a bespoke scheme to provide users with
  increased privacy, achieved through use of \textbf{stealth addresses} to hide
  transaction data, and \textbf{key images} to prevent double spending
  \cite{CryptN}. Key images are the result of a one-way function being
  performed on the given user's private key.

Monero use ring signatures in a `passive' mixing protocol
  \cite{noether2016ring}.  Each transaction is signed using Monero's ring
  signature scheme, which produces a key image, containing information letting
  third parties know that the transaction has been formed correctly and is not
  an attempt to double spend.

Ring signatures are combined with stealth addresses, one time use addresses
which are not associated with any user. The recipient of the coins can then
identify where they are stored by using a private `viewkey'. They can then be
spent by this recipient by him forming a ring signature with his private
`spending key' \cite{CryptN}.

Stealth addresses were suggested by Bitcoin developer Peter Todd, and are
widely used in bitcoin.  The ideas is as follows -- instead of having one
address, which leads to easy tracing, or having many addresses, which means a
user must store and remember multiple different private keys, stealth addresses
allow user to use multiple addresses as if they were just one -- a server
handles the multiple accounts to improve user experience, coin fungibility and
user privacy.

Ring signatures are combined with Greg Maxwell's \textbf{Confidential
Transaction} (CT), forming `Ring Confidential Transactions' \cite{CT}. CTs are
used to hide the value of the transaction, but not the sender or recipient
address.  This is achieved through additively homomorphic encryption, resulting
in transactions around 5kB, and proof size 2.5kB \cite{CT}.

Monero uses an original method in order to hash to an elliptic curve, one that
doesn't not appear in any reseach papers, although the Monero team claim it is
`a secure hash function' \cite{shnoe}.  However, there is no analysis of
whether the function output is distributed uniformly at random, leading to it
being accepted as indifferentiable from a random oracle, and whether the
implementation is truly one way.  Monero have chosen to use an Edwards curve to
base their EC cryptography, due to the higher speed and, under some
definitions, higher level of security offered by Curve25519 \cite{safecurves}.

\subsection{An Anonymous Cryptocurrency Comparison} The cryptocurrencies
discussed above are addressing the privacy problem with different approaches,
and so are appropriate for use in different situations.

For example, ZCash requires a trusted set up stage, but after that the system
is entirely anonymous.  Due to the nature of the system and its use of
zero-knowledge proofs, after the first transaction involving a coin, all coins
are entirely anonymous and the blockchain is `opaque', revealing nothing about
senders, recipients, or transaction values \cite{hopwood}.

Although an essential property for a cryptocurrency with total anonymity, this
opaqueness makes the scheme impossible to audit.  It  is also impossible to see
whether forged coins have entered into the blockchain system, and so parties
involved in the set up process must be trusted to act honestly for all time.

Ring signature mixes (RMSs) do not offer this same level of anonymity, which
results in the underlying blockchain system remaining auditable.  The
`plausible deniability' or reduced level of anonymity provided is sufficient to
make statistical analysis cumbersome and unenlightening, but the scheme as a
whole is still possible to audit as a unit.  RSMs do not require any additional
trust, which lends well to use within cryptocurrency systems.

RSMSs have much faster generation and verification stages than ZCash style
zero-knowledge proofs. ZCash proofs, although fairly small at 322 bytes, and
very fast to verify, take `one to two minutes' to produce \cite{hopwood}.

Linkable ring signatures have applications beyond the blockchain, for example
enabling spam prevention on anonymous online forums or chat rooms, through rate
limiting or blocking known spammer signatures. Another possible use case in
that of random authentication from within a group - for example an attacker
would find it much more difficult to perform a DoS attack or otherwise change
the behaviour of a group of servers if the server requests and responses were
signed by a randomly selected server producing a ring signature, rather than
each server providing authentication by producing a digital signature.

\section{Implementation \& Methodology}

\subsection{Our Mixing Contract at a High Level} Assuming that the blockchain
is equipped with an adequate scripting language, for example Ethereum's
Solidity, or Rootstock for bitcoin \cite{RSK}, the unique ring signature based
mix scheme is implemented as follows:

\begin{enumerate} \item A contract is made to verify ring signatures, receive
      and distribute coins on the blockchain. Parameters for the specific mix
      are entered into the contract.

  \item Users generate public and private elliptic curve key pairs.  These are
    \textbf{not} the public and private keys associated with the accounts that
    the coins are being sent to or from.  They are from an off-chain public key
    infrastructure.  We use randomly generated elliptic curve key-pairs
  (explored in more detail in Section 2.1.1) \item Users wishing to participate
    in the ring mix send their public key and the agreed denomination of the
    cryptocurrency, for example 1 Ether, to the contract.  When a sufficient
    number of users have sent their public keys to the contract, with
    sufficient defined in respect to the original contract parameters, the
    contract publishes the list of public keys which together form the ring.

  \item The intended recipients of the coins from the mix are the holders of
    the secret keys corresponding to the public keys submitted to the contract.
    The recipient can be the same user as the sender, in which case that person
    can generate the key pair, or the recipient can be different to the sender,
    which requires the \emph{recipient} of the funds to generate the key pair,
    and send only the public key to the sender of the funds.

  \item Intended recipients send the signature to the contract. The signature
    includes a tag, which is unique to each signer, message, and ring.

  \item The contract verifies that the tag is formed correctly, corresponding
  to one of the public keys in the ring.  The signature and tag will only
verify if: \begin{itemize} \item The message signed is the correct message,
    \item The ring in question is correct, \item The tag is correctly formed,
\item The tag has not been seen before.  \end{itemize} \item Funds are released
to each sender of a verified signature and tag.  \end{enumerate}

\subsection{The Franklin-Zhang URS Scheme} The unique ring signature scheme is
as follows, with the signer as the $i^{th}$ user in the ring.  We use the
notation $ \leftarrow_R $, to indicate choosing an element at random from a
set, for example $ t_j \leftarrow_R \mathbb{Z}_q $ shows $ t_j $ chosen at
random from $ \mathbb{Z}_q $.  \begin{itemize} \item
      \textbf{Setup($1^{\lambda}$)}: For $\lambda$ the security parameter,
      choose multiplicative group $\mathbb{G}$ with prime order $q$, and
      randomly chosen generator $g$ of $\mathbb{G}$. Choose also two hash
      functions $H$ and $H'$ such that: \begin{itemize} \item $H: \{0,1\}*
      \rightarrow \mathbb{G}$ \item $H': \{0,1\}* \rightarrow \mathbb{Z}_q$
      \end{itemize} Output public parameters \textbf{pp} = ($\lambda$, $q$,
      $\mathbb{G}$, $H$, $H'$).

  \item \textbf{RingGen($1^{\lambda}$, pp)}: Key generation algorithm for user
  $i$: \begin{itemize} \item $x_i \leftarrow_R \mathbb{Z}_q$ \item $y_i
  \leftarrow g^{x_i}$ \end{itemize} Output public key $pk_i$ = (\textbf{pp},
    $y_i$), secret key $sk_i$ = (\textbf{pp}, $x_i$).

  \item \textbf{RingSig($sk_i, R, m$)}: The below is the protocol to sign
  message $m$ in the ring with description $R = (pk_1, \ldots, pk_n)$.
\begin{enumerate} \item For $j \in [n]$, $j \neq i$, we have: \\ $t_j, c_j
      \leftarrow _R \mathbb{Z}_q$, \\ $a_j \leftarrow g^{t_j}y_j^{c_j}$, \\
    $b_j \leftarrow H(m||R)^{t_j}(H(m||R)^{x_i})^{c_j}$.  \item For $j = i$, we
      have: \\ $r_i \leftarrow_R \mathbb{Z}_q$, \\ $a_i \leftarrow
      g^{t_j}y_j^{c_j}$, \\ $b_i \leftarrow H(m||R)^{r_i}$.  \item Calculate
      $c_i \leftarrow [H'(m, R, \{ a_j, b_j \}^n_1) - \sum_{j \neq i}c_j]$ mod
    $q$, \\ $t_i \leftarrow r_i - c_i x_i$ mod $q$.  \item Return $(R, m,
      H(m||R)^{x_i}, c_1, t_1, \ldots, c_n, t_n)$.  \end{enumerate}

  \item \textbf{RingVer($R, m, \sigma$)}:\\ Parsing the output of
    \textbf{RingSig}, and using the notation $H(m||R)^{x_i} =
    \mathlarger{\tau}$, we perform the comparison: $$ \sum^n_1 c_j = H'(m, R,
    \{g^{t_j}y_j^{c_j}, H(m||R)^{t_j} \mathlarger{\tau}^{c_j} \}^n_1). $$
\end{itemize}

There is an important line given in this scheme, with important implications to
note that when implementing this scheme over ECs.  Starting from an assignment
given in the scheme, we have: \begin{align*} & t_i \leftarrow r_i - c_i x_i
  \text{ mod } q, \implies & r_i = t_i + c_i x_i \implies & g^{r_i} = g^{t_i}
g^{c_i x_i} \implies & g^{r_i} = g^{t_i} (g^{x_i})^{c_i} \implies & g^{r_i} =
g^{t_i} y_i^{c_i}.  \end{align*}

Although these implications hold unconditionally over the integers, when
defined over elliptic curves we much instead define them modulo the order of
the \textbf{generator} of the EC in question, rather than the generator of the
base field.

In our implementation, the generator order is labelled $n$ -- when instead, the
exponents of the generator are expressed modulo $q$, equalities such as $g^{c_i
x_i} = (g^{x_i})^{c_i}$, which are essential for correct, pubic verification of
the scheme, as the precomputed $g^{x_i}$ is a public constant, but $x_i$ itself
is not.

\subsubsection{The Franklin-Zhang URS Construction} Franklin and Zhang's unique
ring signature scheme relies on Blum, Feldman and Micali's transformation to
the original zero knowledge proof scheme \cite{f2012}.  Blum, Feldman, and
Micali (BFM) applied a transformation to the original zero-knowledge proof
scheme, producing a \textbf{non-interactive} zero-knowledge proof (NIZK)
scheme, improving greatly on the efficiency of the formerly interactive scheme
\cite{blum1988non}.

With a hash function $H(\cdot)$ modelled as a random oracle, it is a known
result that setting $F = H^x$ constructs a pseudo-random function $F$.  As the
security of our unique ring signature scheme relies on the `random oracle'
property of this hash function, we must be careful to use a
\textbf{well-distributed} hash function.

Proof of membership is explained most simply as a proof of knowledge of (at
least) one of a specific set of numbers. The notable change is that rather than
constructing a random challenge, the challenge takes into account the
randomness that was incorporated the scheme.  In the case of the URS, this
randomness corresponds to each of the other public keys in the ring.  is
included in the scheme as a way of `blinding' the true signer, as
probabilistically, each of the numbers are equally likely to have been
generated at random.

In our scheme, we have: \begin{align} a_i &= g^{r_i} = g^{t_i}y^{c_i} =
g^{t_i}g^{x_ic_i} \\ b_i &= H(m||R)^{r_i} = H(m||R)^{t_i}(H(m||R)^{x_i})^{c_i}
\end{align} The exponent $t_i$ is applied to $H(m||R)$, which acts as a random
oracle and is uniformly distributed in accordance with Franklin and Zhang.  The
scheme is congruent to a typical zero knowledge proof, with the following
substitutions: \begin{itemize} \item Random witness: $W = g^{r_i}$.  \item
      Random challenge: $c_i = H'(m, R, \{a_j, b_j\}_1^n ) - \Sigma c_j \text{
        mod } q$.  \item Response: With $t_i = r_i - c_i x_i \text{ mod } q $,
      $a_i$ and $b_i$ are formed as shown in equations (5) and (6) above.
    \item Verification: $a_i = g^{r_i} = g^{t_i}y^{c_i} = g^{t_i}g^{c_i x_i}$
      and $b_i = H(m||R)^{r_i}$, similarly.  \end{itemize} We use the two
      generators to ensure that the equality holds for both the specific ring
      and message in question, and in the general case with the group
      generator.

An explanation of this proof is given in Appendix D, and is followed by proofs
of completeness, unforgeability and anonymity, which are give in Appendices D,
E, and F, respectively.  The proofs of unforgeability and anonymity are
explored through games, and adapted from the proofs given in the Franklin and
Zhang paper \cite{f2012}.  The proof of completeness and the exploration of the
NIZK scheme used are new, but were constructed in a straight-forward manner
from the URS scheme.

It is also important to note here that as we are constructing these signatures
within EC groups, there are some technicalities which must be addressed for the
scheme to work correctly.

\subsection{Hashing to Elliptic Curves}

The Franklin-Zhang URS scheme requires us to construct a hash of the form: $$ H
\colon \{ 0, 1 \} ^* \to \mathbb{G}. $$

We have chosen to implement the system over EC groups for both efficiency of
implementation and security, rather than the alternative of multiplicative
groups formed by finite prime fields.  Hence, we require a hash function of the
form: $$ H \colon \{ 0, 1 \} ^* \to E(\mathbb{F}_q). $$ More specifically, we
have chosen to implement the scheme over the EC used in both bitcoin and
Ethereum, secp256k1, which is defined in Section \ref{bitcoin}.

There are several algorithms that can potentially be used to hash from an
arbitrary length string to an EC group.  The simplest methods are ``try and
increment'', and simply using an existing, secure hash function, and
multiplying the output by a generator of the EC group in question. Both of
these options are explored below.  Other hashing to EC algorithms can be placed
into two broad categories: ``Icart-like'' functions, based on the algorithms
proposed by Icart in his 2009 `How to Hash into Elliptic Curves'
\cite{HashtoEC}, and the more general Shallue-Woestijne-Ulas (SWU) algorithm,
proposed in 2006 in the paper titled `Construction of rational points on
elliptic curves over finite fields' \cite{shallue}.  In order for either scheme
to produce a reasonable output, certain criteria must be met by the underlying
field over which the EC group in question is defined. Some common EC choices
and their parameters are displayed in Table \ref{comp}.

\begin{table}[h] \centering \caption{A comparison of some common EC
  parameters.\label{comp}} \begin{tabular}{ c c c c }\toprule Source & EC &
    Security & Parameters \\ \midrule CNSS 2012 & p256 & 128 & a $= -3$, $q =
    2^{256} - 2^{224} + 2^{192} + 2^{96} - 1 $ \\ CNSS 2015 & p384 & 192 & a $=
    -3$, $q = 2^{384} - 2^{128} - 2^{96} + 2^{32} - 1$ \\ CNSS 2015 & p521 &
    260  & a $= -3$, $q = 2^{384} - 2^{128} - 2^{96} + 2^{32} - 1$ \\ bitcoin &
    secp256k1 & $\approx$ 128 & a $= 0$, b $= 7$, $q =  2^{256} - 2^{32} - 977
    $ \\ Bernstein & Curve25519 & $ \approx$ 128 & $ y^2 = x^3 + 486662x^2 +
  x$, $q = 2^{255} - 19$ \\ \bottomrule \end{tabular} \end{table}

\subsubsection{secp256k1}\label{bitcoin}

Bitcoin and Ethereum both use the Koblitz curve \textbf{secp256k1}, recommended
  by SECG \footnote{SECG:\ Standards for Efficient Cryptography Group} in 2000
  \cite{certicom2000sec}, over which to perform ECDSA to produce the signatures
  on transactions \cite{bit}.  secp256k1 was chosen over the random secp256r1
  in order to avoid having to trust the `randomness' used to generate the
  curves parameters, and to avoid the possibility of a backdoor being included
  in this randomness.  Koblitz curves have a particular structure which allows
  very fast performance when implementing EC point addition and multiplication
  by a scalar \cite{gallant2001faster}.

The increased speed in implementation of general elliptic curve arithmetic over
  secp256k1 also corresponds to increased efficiency of the fastest known
  attacks on the ECDLP.  As a result of this, secp256k1 is `several bits' less
  secure than one would normally expect a 256 bit elliptic curve to be (which
  is $\sim$128) \cite{safecurves}.  The use of secp256k1 to secure all money in
  both the bitcoin and Ethereum systems has drawn criticism from some
  researchers \cite{nicslide}, but generally the elliptic curve choice is not
  thought of as weak.

secp256k1 consists of all the rational points $(x, y) \in E(\mathbb{F}_q)$ that
  satisfy\footnote{A brief overview of the EC cryptography leading up to how
  this equation is formed is given in Appendix A.}.  \begin{equation} y^2 = x^3
  + 7.  \end{equation}

\subsubsection{Simple Methods}

\subparagraph{Generator Multiplication} Shadowcash is an example of a
  cryptocurrency that tried to use this straight-forward but naive approach to
  `hash' to an elliptic curve point. With $G$ a group generator, ShadowCash
  achieves an EC hash output through the construction \cite{shnoe}: $$ H =
  \text{SHA3}(y_i) \cdot G.$$

As usual, we have that the public key is formed $y_i = x_i \cdot G$, with $x_i$
  the randomly generated secret key.  Although this seems like a reasonable
  approach to producing random EC points, using this method removes all privacy
  enhancing properties of the scheme, as identifying information can be
  revealed without the discrete logarithm problem being solved.  Instead, the
  public key of the signer is trivially revealed, allowing the input and output
  addresses to be linked in the same way as a transaction not involving ring
  signatures.

ShadowCash uses key images in place of Franklin-Zhang's tags, although they are
  constructed very similarly -- a ShadowCash key image, $\mathlarger{\tau}$, is
  formed: $$  \mathlarger{\tau}  = x_i \cdot H_p(y_i).$$

Rearranging this equation and including the construction of the hash function
  $H_p$: \begin{align*} \mathlarger{\tau} & = x_i \cdot \text{SHA3}(y_i) \cdot
  G \\ & = \text{SHA3}(y_i) \cdot (x_i \cdot G) \\ & = \text{SHA3}(y_i) \cdot
  y_i.  \end{align*} It is clear that $\mathlarger{\tau}$ can be produced using
  only public information -- all we require is access to the SHA3 function, and
  the public key of each individual in the ring.  We can produce
  $\mathlarger{\tau}$ corresponding to each public key and compare against the
  tag in each signature, resulting in the signers having only the pseudonymity
  they achieve using bitcoin or a similarly unblinded cryptocurrency.

In our case, the identification tag is $H(m||R)^{x_i}$, which corresponds to
$x_i \cdot H(m||R)$ when defined over an elliptic curve group. An analogous
attack could be carried out by an adversary with knowledge of the message being
signed, and all public keys belonging to the ring.

\subparagraph{Try and Increment} This method is applicable to all ECs, however,
is probabilistic, so vulnerable to timing based side channel attacks and will
not find a point with certainty.  From Icart's `How to Hash into Elliptic
Curves', with security parameter $k$, the try and increment scheme is given by
Algorithm 1.

\begin{algorithm} \caption{Try and Increment} \begin{algorithmic} \STATE Input
  $u$.  \FOR{$i = 0$ to $k - 1$} \STATE $x = u + i$ \IF{ $x^3 + ax + b$ is a
  quadratic residue in $\mathbb{F}_q$} \RETURN $Q = (x, (x^3 + ax + b)^{1/2})$
  \ENDIF \ENDFOR \RETURN $\bot$ \end{algorithmic} \end{algorithm}

We can use the try and increment method without compromising the security of
our scheme as the input information does not need to be kept secret. So timing
or other side channel attacks made possible through the non-constant time
function would reveal nothing besides public information to the adversary.

There are several well-tested schemes for finding rational points on elliptic
curves, such as \cite{shallue}, \cite{poonen2001computing},
\cite{fisher2008finding}.  However, as these schemes are not for `hashing' to
an elliptic curve, there is no consideration given to the distribution of the
points produced, and the `indifferentiability' of the scheme compared to with
the random oracle model, which is essential for our scheme.

\subsubsection{Schemes from Literature}

As mentioned above, EC parameters generally have to satisfy some criteria
before a known algorithm can be used with provable security guarantees.
Icart-like functions build hash functions of the form \cite{HashtoEC}: $$ H(m)
= f(H'(m)),$$ with $H'$ modelled as a random oracle, and practically
implemented with an existing, secure hash function, and $f$ a hash encoding
first suggested, and detailed in full, in Icart's `How to Hash into Elliptic
Curves' \cite{HashtoEC}.
These actually cannot be used within our specific scheme, as Ethereum's Virtual
Machine (the EVM) must be purely deterministic, and so the randomness used
within these functions is unavailable.
Nevertheless, we will review possibilities in case of future relevance.

With $q$ the order of the underlying field for the elliptic curve (so the
$\mathbb{F}_q$ used to construct $E(\mathbb{F}_q)$), Icart's function requires
both that $q \equiv 3 \text{ mod } 4$ \cite{HashtoEC}, and also that $q \equiv
2 \text{ mod }3$.  secp256k1 satisfies the former condition, but has $q \equiv
1 \text{ mod } 3$, and so neither of these schemes are suitable.

The SWU scheme again requires $q \equiv 3 \text{ mod } 4$, which secp256k1
satisfies, but also, for an EC defined $y^2 = x^3 + ax + b$, the scheme
requires that $a, b \neq 0$.  As secp256k1 is defined with equation $y^2 = x^3
+ 7$, $a = 0$, and this function cannot be used.

Another scheme, suggested by Brier et al., \ takes the form $H(m) = f(h_1(m)) +
f(h_2(m))$ with $h_1$, $h_2$ independent random oracles with values in
$\mathbb{F}_q$, $f$ as Icart's encoding \cite{brier2010efficient}.  However,
this also requires $q \equiv 2 \text{ mod }3$.

A comparison of some typical choices of ECs, and whether they satisfy the
necessary criteria, is shown in Table \ref{req} below.

\begin{table} \centering \caption{A comparison of several ECs, with typical
  hash-to-EC function requirements.\label{req}} \begin{tabular}{c c c
    c}\toprule Curve & q mod 3 & q mod 4 & q mod 12 \\ \midrule p224 & 1 & 1 &
    1 \\ p256 & 1 & 3 & 7 \\ p384 & 2 & 3 & 11 \\ p521 & 1 & 3 & 7 \\ secp256k1
  & 1 & 3 & 7 \\ \bottomrule \end{tabular} \end{table} In elliptic curve groups
  where the equivalence $q \equiv 3 \text{ mod } 4$ holds, square roots modulo
  $q$ can be calculated efficiently and deterministically through use of the
  Tonelli-Shanks algorithm, given below. A critical review of this algorithm
  can be found in \cite{lindhurst1999analysis}, and notably the algorithm does
  not run in constant time, and rather the timing follows a normal
  distribution.

\begin{defn}{Tonelli-Shanks Algorithm} For $q$ an odd prime, with $q \equiv 3
  \text{ mod } 4$, and $n \in \mathbb{F}_q$, the two square roots of $n$ are
  $u$ and $-u$, with $u$ calculated as given: $$ u = n^{\frac{q+1}{4}} \text{
    mod } q.$$ \end{defn}

This can be shown to hold through a very straight-forward deduction, starting
with Euler's criterion.  \begin{defn}{Euler's Criterion} \begin{enumerate}
  \item If $n$ is a square modulo $q$, then $n^{\frac{q-1}{2}} \equiv 1 \text{
      mod } q$.  \item If $n$ is not square modulo $q$, then $n^{\frac{q-1}{2}}
\equiv -1 \text{ mod } q$.  \end{enumerate} \end{defn}

Another scheme, suggested by Brier et al., \ takes the form $H(m) = f(h_1(m)) +
  f(h_2(m))$ with $h_1$, $h_2$ independent random oracles with values in
  $\mathbb{F}_q$, $f$ as Icart's encoding \cite{brier2010efficient}.  However,
  this also requires $q \equiv \text{2 mod 3}$.

\subsection{secp256k1 -- not quite a Barreto-Naerhig curve} The anonymity
  property of the URS scheme we have implemented relies on the hardness of the
  DDH problem in the group we are working in \cite{f2012}.  Due to this, if
  there exists a group pairing, from secp256k1 to a field in which the DDH
  problem is trivial to break, even the weakest adversaries would have the
  ability to remove the additional anonymity given by the ring signature based
  mixing scheme.

A \textbf{pairing-friendly} curve is one with an embedding degree that is very
  low, which leads to the finite field being mapped to having a small size, and
  hence low bitwise security.  The embedding factor is the factor by which the
  storage costs grow when storing the field that the elliptic curve has been
  mapped to. With embedding degree $d$, the field size (of the new field in
  which we can break the DDH problem) is $2^{256 \cdot d}$.

With the underlying finite field $\mathbb{F}_q$ defined such that $q \equiv 1
  \text{ mod } 3$, Barreto-Naehrig curves are defined with the equation: $$ y^2
  = x^3 + b, \qquad \qquad \qquad b \in \mathbb{F}_q.$$ It is trivial to see
  that secp256k1 has this form, and it can be shown that the condition on $q$
  is also satisfied, as is given in Table \ref{req}.

However, Barreto-Naerhing curves have an embedding degree of $d = 12$
  \cite{barreto2005pairing}, and secp256k1 specifically has an embedding degree
  of \cite{safecurves}: \begin{multline} d = 19298 \ 68153955 \ 26992372 \
    61830834 \ 78131797 \\ 54729273 \ 79845817 \ 39710086 \ 05235863 \
    60249056.  \end{multline} This means a single pairing value would require
    $256 \cdot d$ bits of storage -- approximately $5 \times 10^{78}$. Even
    with the more optimal $d = 12$, the finite field with which the elliptic
    curve group is paired grows to an extent that the discrete logarithm
    problem in the finite field requires more computational time than Pollard's
    Rho in the elliptic curve group \cite{barreto2005pairing}.  Clearly, a
    pairing-based ECDLP break is not a threat at present.

\subsection{Hashing to Barreto-Naehrig Curves} The scheme we choose to
implement works as following. Although technically not a Barreto-Naehrig curve,
secp256k1 satisfies all of the requirements for this hashing to elliptic curve
algorithm to work.  This scheme is elegant, deterministic and executes in
constant time, eliminating the possibility of side channel attacks.

The scheme used and its security properties are given in full in Fouque and
Tibouchi's `Indifferentiable Hashing to Barreto-Naehrig Curves' \cite{BNCurve}.
Importantly, it is proven that the scheme is indifferentiable from a random
oracle.  With $\#E(\mathbb{F}_q)$ defined as the number of rational points on
the elliptic curve in question, the image of the hashing function is
approximately $\frac{9}{16} \cdot \#E(\mathbb{F}_q)$. For comparison, Icart's
function produces an image of $\frac{5}{8} \cdot \# E (\mathbb{F}_q )$ if $ a
\neq 0$, and $\frac{2}{3} \cdot \# E (\mathbb{F}_q)$ if $ a = 0 $
\cite{Giechaskiel:2016wv}.

As we need to use the hashing function to generate the tags and the
zero-knowledge proof of membership, as constructed $H(m||R)^{w}$, with $w =
x_i, r_i, t_i$, we need the hash to act as a random oracle in order for the
$H^w$ to act as pseudo-random functions as defined in the Franklin and Zhang
paper \cite{f2012}.  The hashing scheme we have implemented is shown in
Algorithm 2, with $t$ as input, and $b$ as defined in the EC equation $y^2 =
x^3 + b$, with $(x, y) \in \mathbb{F}_q \times \mathbb{F}_q$, taken from
\cite{BNCurve}:

  \begin{algorithm}[H] \caption{Hashing to a Barreto-Naehrig Curve}
    \begin{algorithmic} \STATE input $t$ \STATE $ w \leftarrow \frac{\sqrt{3}
      \cdot t}{1 + b + t^2}$ \STATE $ x_1 \leftarrow \frac{-1 + \sqrt{-3}}{2} -
      tw$ \STATE $ x_2 \leftarrow -1 - x_1$ \STATE $ x_3 \leftarrow 1 +
      \frac{1}{w^2}$ \STATE $ r_1, \ r_2, \ r_3 \leftarrow_R \mathbb{F}^*_q $
      \STATE $ \alpha \leftarrow \chi_q(r_1^2 \cdot (x_1^3 + b))$ \STATE $
      \beta \leftarrow \chi_q(r_2^2 \cdot (x_2^3 + b))$ \STATE $ i \leftarrow [
    ( \alpha - 1 ) \cdot \beta \text{ mod } 3 ] + 1$ \STATE return $ ( x_i,
    \chi_q(r_3^2 \cdot t) \cdot \sqrt{x_1^3 + b} )$.  \end{algorithmic}
  \end{algorithm}

$ \mathlarger{\chi} $ is a function defined in the Foque and Tibouchi paper
such that: \begin{equation*} \mathlarger{\chi(a)} = \begin{cases} 0, & \text{if
} a = 0, \\ 1, & \text{if } a \text{ is a square}, \\ -1, & \text{otherwise}.
\end{cases} \end{equation*}

 We construct the $\mathlarger{\chi}_q$ function for use within the hashing
 algorithm using Euler's Criterion as previous defined, with $q$ the order of
 the desired output EC group. This is possible due to secp256k1 satisfying $q
 \equiv 3 \text{ mod } 4$, as explained previously.

Using this scheme as $H$, and SHA-256 as $H'$, we can very easily and
efficiently implement the full unique ring signature algorithm, using only
either Ethereum or bitcoin's existing dependencies.  \section{Evaluation}

\subsection{Privacy} There is a real risk of all but one of the participating
parties in the mixing contract, either through a Sybil attack or collusion,
revealing their tags or secret keys, and so eliminating themselves from the
anonymity group, and therefore linking the honest user's input and output
addresses.

There are several ways to mitigate against this. Chaum blinding could be used
to prevent this attack from occurring, or we could punish users who reveal
their signatures, in order to keep the scheme auditable.  The more suitable
option would be decided based on the use case -- for example, in a financial
environment or in a consortium use case, simply punishing users may be the
easier and preferable option.  Alternatively, we could randomly allocate the
mix that users get entered into, which would make coordinating such a
de-anonymising attack much more difficult.

Otherwise, the privacy of the scheme relies on ECDDH, which itself is closely
related to ECDLP, which has been described as `the hardest math problem ever'
\cite{petit2016algebraic}.  Although currently less efficient than even a brute
force attack, index calculus style attacks are being developed against certain
EC groups, for example p-224. As p-224 also does not satisfy the requirements
for any known hashing to EC function that is indifferentiable from a random
oracle, it is not a wise choice for use in this scheme.

Although transactions can be split across multiple mixes and the total
transaction value can be obfuscated in this way, we have not directly
incorporated hiding of transaction value within our scheme.

\subsection{Scheme Security}

The scheme we have implemented is secure under the EC-DDH assumption in the
random oracle model. Before proving soundness, completeness and anonymity, we
must formally define them in relation to our scheme.  Franklin and Zhang start
the proof that the zero-knowledge proof scheme is sound, but do not finish the
proof or extend to the unique ring signature protocol.

The authors specifically say that it is more efficient in both the signing and
verification algorithms than those suggested in `Linkable Spontaneous Anonymous
Group Signatures in Ad Hoc Groups', by Liu, Wei, and Wong [2004].

Security is generally described through `games' - interactions between a
challenger and a computationally bound adversary, with restrictions based on
the threat model of the scheme. This is how the security level of the protocol
is produced, as shown by the security proofs detailed in Appendices D, E, and
F.

\subsection{Efficiency} Although the scheme is noted as being sub-linear in the
number of users in the ring, it holds rather that the \textbf{tag} is
sub-linear (in fact, constant size), but the full signature, as we have seen
before, is constructed as: \begin{align*} \mathlarger{\sigma} = & \
\mathlarger{\tau}, \ m, \ R, \ \{c_i, t_i \}^n_1 \\ = & \ \mathlarger{\tau_x},
\ \mathlarger{\tau_y}, \ m, \ R, \{c_i, t_i \}^n_1 \end{align*}

Depending on the construction of the contract which uses this ring signature
scheme, it may be possible to simply send $\mathlarger{\tau}$ and $ \{ c_i, t_i
\}^n_1 $. However, as each $c_i$ and $t_i$ are 32 bytes, and taking into
account $\mathlarger{\tau}$, this signature is still $64(n + 1)$ bytes in
length.  Contrasting with ZCash, where each proof is 322 bytes, we achieve a
shorter signature length for rings with size $n < 4 $, but a zk-SNARK is
smaller for $ n \ge 5 $.

However, ZCash is currently running on a `test-net' only, and the fully
functional version is due for release in October 2016 \cite{hopwood}. Our ring
signature scheme, although expensive to implement on Ethereum, is ready to be
used.  The compromise between anonymity guarantees and scheme usability is at
the discretion of the user, but zk-SNARKs would required a large improvement in
key and proof generations times until they are comparable to URS-based mixing
schemes in this aspect.

\subsection{Usability}

It may be possible to implement the scheme in a more user-friendly manner in
the future. For example, more of the mixing contract may be automated, for
example setting up rings in an ad-hoc way, and removing the necessity of sender
and recipient interaction.

For transactions of large value or if there is not a lot of liquidity in the
scheme, we could dynamically adjust the timings of the coin mixes so as to
increase anonymity as much as possible.  For example, if a large transaction is
sent, rather than hoping there are other transactions of the same value for the
scheme to mix among, we will split the transaction into denominations with high
liquidity and mix.  We may be able to further improve the anonymity of this
scheme by not including all of these segments of the transaction in the same
block, unlinking the input and output addresses of the full, larger transaction
so as to make the large transaction less conspicuous even under careful
blockchain analysis.  We could do this automatically, although we would need to
implement the scheme in a completely different way, as the senders and
recipients may no longer be able to interact with just one contract.

Another idea for consideration, if sub-linear linkable ring signatures were
available, would be to include all previously used public keys in the ring. We
could do this up to a predefined maximum, and select the public keys to be used
at random, so that analysis the ring of one transaction and the ring of the
previous transaction would not give away the second signer, as theirs would be
the only public key not common to the two sets.  We could also look into
hashing the set of public keys, rather than appending them in the signature, to
reduce signature size.

\section{Conclusions, Summary, Further Work}

\subsection{Conclusions} This research has examined the existing solutions to
the lack of inherent privacy for individuals in current blockchain systems.
The research showed that there are schemes that offer anonymity on the
blockchain, and schemes that offer a form of anonymity that we refer to as
plausible deniability.  Both of these have their possible limitations -- for
the truly anonymous blockchain systems, there is an `opaqueness' to the once
transparent ledger of transactions, and their is additional trust required to
construct such a scheme. For the platforms offering `plausible deniability',
there are often vulnerabilities introduced to the scheme privacy due to
incorrect implementation or practical limitations such as small anonymity
groups, and there are ways for a motivated adversary to subvert the privacy of
existing schemes functioning in this way.

There are no schemes or systems currently available or proposed that enable a
blockchain platform to retain some of its transparency, while enabling
individual users to remain truly anonymous.

\subsection{Summary} We have explored cryptographic protocols, specifically
concentrating on Franklin and Zhang's unique ring signature protocol. We have
implemented this scheme over secp256k1, enabling compatibility with both
bitcoin and Ethereum's EC libraries.

We have clarified an essential condition for a URS to be constructed correctly
over an EC group - all variables given in the Franklin-Zhang algorithm that are
intended to be used as exponents, or rather as a scalar multiplier of a point
in the EC group, must be defined modulo the EC group generator order, rather
than the order of underlying finite field.

We have evaluated the security of the scheme, the privacy offered by the
scheme, and the common vulnerabilities introduced by imperfect implementations
of similar protocols.

\subsection{Further Work} There are many open problems in this general area,
ranging from constructing a blinding system that works alongside linkable or
unique ring signatures, to making existing linkable ring signature schemes more
efficient and scalable.  Sub-linear unique ring signatures exist, although they
are based on bilinear pairings of ECs, and so implementing them is a much
larger project than this one.

We could use a tree structure to produce a proof of knowledge of one public key
of the many potentially contained on the leaves of the tree -- this structure
would allow proofs to be verified in logarithmic space and time
\cite{szydlo2004merkle}, and can be instantiated as a zero-knowledge proof.

There are large improvements to be made in the efficiency of the Ethereum
Virtual Machine, where our final scheme will be implemented. At present, it has
`precompiles' for verifying ECDSA, querying RIPE160 and SHA256, but there are
only basic implementations of arbitrary precision big integer arithmetic, and
no elliptic curve arithmetic other than ECDSA verification.

We could also approach the privacy problem with zk-SNARKs, exploring options to
implement them in such a way that the blockchain platform as a whole remains a
publicly verifiable system.  As each potential solution is made up of many
cryptographic protocols, there are many areas where efficiencies can still be
gained, or security of schemes can be improved.

\newpage
\bibliography{bib} \bibliographystyle{plain}

\begin{thebibliography}{10}

\bibitem{FPGA}
{Custom FPGA Board for Sale!}
\newblock \url{https://bitcointalk.org/index.php?topic=37904.0}, August 2011.
\newblock {Accessed: 2016-10-02}.

\bibitem{sharedcoinbreak2}
{Kristov Atlas Security Advisory}.
\newblock \url{http://www.coinjoinsudoku.com/advisory/}, June 2014.
\newblock {Accessed: 2016-08-19}.

\bibitem{SDC}
{[SDC] ShadowCash | POSV3 | Untraceable E-Cash | NIZKP | HD+BIP32 |
  ShadowMarket*}.
\newblock \url{https://bitcointalk.org/index.php?topic=745352.0}, August 2014.
\newblock {Accessed: 2016-08-16}.

\bibitem{sharedcoinbreak}
{Shared Coin Suspended Temporarily}.
\newblock
  \url{https://blog.blockchain.com/2014/03/17/shared-coin-suspended-temporarily},
  March 2014.
\newblock {Accessed: 2016-08-19}.

\bibitem{jm}
{[ANN] Joinmarket - Coinjoin that people will actually use}.
\newblock \url{https://bitcointalk.org/index.php?topic=919116.0}, January 2015.
\newblock {Accessed: 2016-08-19}.

\bibitem{ccs}
{Colored Coins - Open source standard for digital assets on top of the bitcoin
  network}.
\newblock \url{http://coloredcoins.org/}, June 2015.
\newblock {Accessed: 2016-08-30}.

\bibitem{bitdevel}
{Developer Guide - Bitcoin}.
\newblock \url{https://bitcoin.org/en/developer-guide#block-chain}, September
  2015.
\newblock {Accessed: 2016-08-18}.

\bibitem{cmc}
{All Currencies - Crypto-Currency Market Capitalizations}.
\newblock \url{coinmarketcap.com/currencies/views/all/}, 2016.
\newblock {Accessed: 2016-07-22}.

\bibitem{bci}
{Bitcoin Charts - Blockchain.info}.
\newblock \url{https://blockchain.info/charts}, August 2016.
\newblock {Accessed: 2016-08-18}.

\bibitem{jmb}
{JoinMarket's privacy is degraded (for a while)}.
\newblock
  \url{https://gist.github.com/chris-belcher/00255ecfe1bc4984fcf7c65e25aa8b4b},
  June 2016.
\newblock {Accessed: 2016-08-19}.

\bibitem{VISA}
{VISA USA Small Business Retail}.
\newblock
  {https://usa.visa.com/run-your-business/small-business-tools/retail.html},
  2016.
\newblock {Accessed: 2016-08-20}.

\bibitem{NXT}
{Whitepaper:NXT}.
\newblock \url{http://wiki.nxtcrypto.org/wiki/Whitepaper:Nxt}, February 2016.
\newblock {Accessed: 2016-10-02}.

\bibitem{BTCUserPriv}
Elli Androulaki, Ghassan~O Karame, Marc Roeschlin, Tobias Scherer, and Srdjan
  Capkun.
\newblock {Evaluating User Privacy in Bitcoin}.
\newblock In {\em International Conference on Financial Cryptography and Data
  Security}, pages 34--51. Springer, 2013.

\bibitem{bitsecure}
Marcin Andrychowicz, Stefan Dziembowski, Daniel Malinowski, and Lukasz Mazurek.
\newblock Secure multiparty computations on bitcoin.
\newblock In {\em 2014 IEEE Symposium on Security and Privacy}, pages 443--458.
  IEEE, 2014.

\bibitem{ant2014}
Andreas~M Antonopoulos.
\newblock {\em {Mastering Bitcoin: Unlocking Digital Cryptocurrencies}}.
\newblock O'Reilly Media, Inc., 2014.

\bibitem{SLRSR}
Man~Ho Au, Sherman~SM Chow, Willy Susilo, and Patrick~P Tsang.
\newblock {Short Linkable Ring Signatures Revisited}.
\newblock In {\em Public Key Infrastructure}, pages 101--115. Springer, 2006.

\bibitem{Back:1997}
Adam Back.
\newblock {Hashcash - A Denial of Service Counter-Measure}.
\newblock May 1997.
\newblock {Accessed: 2016-06-13}.

\bibitem{bald2015}
Foteini Baldimtsi, Melissa Chase, Georg Fuchsbauer, and Markulf Kohlweiss.
\newblock {Anonymous Transferable e-cash}.
\newblock In {\em IACR International Workshop on Public Key Cryptography},
  pages 101--124. Springer, 2015.

\bibitem{barreto2005pairing}
Paulo~SLM Barreto and Michael Naehrig.
\newblock Pairing-friendly elliptic curves of prime order.
\newblock In {\em International Workshop on Selected Areas in Cryptography},
  pages 319--331. Springer, 2005.

\bibitem{ZeroCash}
Eli Ben-Sasson, Alessandro Chiesa, Christina Garman, Matthew Green, Ian Miers,
  Eran Tromer, and Madars Virza.
\newblock {Zerocash: Decentralized Anonymous Payments from Bitcoin}.
\newblock In {\em Security and Privacy (SP), 2014 IEEE Symposium on}, pages
  459--474. IEEE, 2014.

\bibitem{ben2013snarks}
Eli Ben-Sasson, Alessandro Chiesa, Daniel Genkin, Eran Tromer, and Madars
  Virza.
\newblock Snarks for c: Verifying program executions succinctly and in zero
  knowledge.
\newblock In {\em Advances in Cryptology--CRYPTO 2013}, pages 90--108.
  Springer, 2013.

\bibitem{ben2015secure}
Eli Ben-Sasson, Alessandro Chiesa, Matthew Green, Eran Tromer, and Madars
  Virza.
\newblock Secure sampling of public parameters for succinct zero knowledge
  proofs.
\newblock In {\em 2015 IEEE Symposium on Security and Privacy}, pages 287--304.
  IEEE, 2015.

\bibitem{cryptoeprint:2014:595}
Eli Ben-Sasson, Alessandro Chiesa, Eran Tromer, and Madars Virza.
\newblock {Scalable Zero Knowledge via Cycles of Elliptic Curves}.
\newblock Cryptology ePrint Archive, Report 2014/595, 2014.
\newblock \url{http://eprint.iacr.org/2014/595}.

\bibitem{BKM}
Adam Bender, Jonathan Katz, and Ruggero Morselli.
\newblock {Ring signatures: Stronger Definitions, and Constructions without
  Random Oracles}.
\newblock In {\em Theory of Cryptography}, pages 60--79. Springer, 2006.

\bibitem{safecurves}
Daniel~J. Bernstein.
\newblock {SafeCurves: Choosing Safe curves for Elliptic-Curve Cryptography}.
\newblock \url{https://safecurves.cr.yp.to/}, January 2014.
\newblock {Accessed: 2016-08-08}.

\bibitem{blum1988non}
Manuel Blum, Paul Feldman, and Silvio Micali.
\newblock {Non-interactive Zero-knowledge and its Applications}.
\newblock In {\em Proceedings of the twentieth annual ACM symposium on Theory
  of computing}, pages 103--112. ACM, 1988.

\bibitem{boneh2005collusion}
Dan Boneh, Craig Gentry, and Brent Waters.
\newblock Collusion resistant broadcast encryption with short ciphertexts and
  private keys.
\newblock In {\em Annual International Cryptology Conference}, pages 258--275.
  Springer, 2005.

\bibitem{bradbury2013problem}
Danny Bradbury.
\newblock The problem with bitcoin.
\newblock {\em Computer Fraud \& Security}, 2013(11):5--8, 2013.

\bibitem{brier2010efficient}
Eric Brier, Jean-S{\'e}bastien Coron, Thomas Icart, David Madore, Hugues
  Randriam, and Mehdi Tibouchi.
\newblock Efficient indifferentiable hashing into ordinary elliptic curves.
\newblock In {\em Annual Cryptology Conference}, pages 237--254. Springer,
  2010.

\bibitem{Buterin:2015d}
Vitalik Buterin.
\newblock {Understanding Serenity, Part 2: Casper}.
\newblock Published at
  \url{https://blog.ethereum.org/2015/12/28/understanding-serenity-part-2-casper/},
  December 2015.
\newblock {Accessed: 2016-04-20}.

\bibitem{Eth}
Vitalik Buterin.
\newblock {Ethereum Whitepaper}.
\newblock Published at \url{https://github.com/ethereum/wiki/wiki/White-Paper},
  March 2016.
\newblock {Accessed: 2016-03-30}.

\bibitem{DAO}
Vitalik Buterin.
\newblock {Hard Fork Completed}.
\newblock \url{https://blog.ethereum.org/2016/07/20/hard-fork-completed/}, July
  2016.
\newblock {Accessed: 2016-08-23}.

\bibitem{DAA}
Jan Camenisch, Ernie Brickell, Liqun Chen, Manu Drivers, and Anja Lehmann.
\newblock {Direct Anonymous Attestation Revisited}.
\newblock
  \url{http://researcher.watson.ibm.com/researcher/files/zurich-JCA/2016-04-14-Direct-Anonymous-Attestation.pdf},
  April 2016.
\newblock {Accessed: 2016-10-02}.

\bibitem{camenisch1999}
Jan Camenisch, Markus Michels, et~al.
\newblock {Separability and Efficiency for Generic Group Signature Schemes}.
\newblock In {\em Annual International Cryptology Conference}, pages 413--430.
  Springer, 1999.

\bibitem{certicom2000sec}
SEC Certicom.
\newblock {SEC 2: Recommended Elliptic Curve Domain Parameters}.
\newblock {\em Proceeding of Standards for Efficient Cryptography, Version}, 1,
  2000.

\bibitem{jens}
Nishanth Chandran, Jens Groth, and Amit Sahai.
\newblock {Ring Signatures of Sub-linear Size without Random Oracles}.
\newblock In {\em Automata, Languages and Programming}, pages 423--434.
  Springer, 2007.

\bibitem{chaum1983blind}
David Chaum.
\newblock Blind signatures for untraceable payments.
\newblock In {\em Advances in cryptology}, pages 199--203. Springer, 1983.

\bibitem{GSig}
David Chaum and Eug{\`e}ne Van~Heyst.
\newblock {Group Signatures}.
\newblock In {\em Workshop on the Theory and Application of of Cryptographic
  Techniques}, pages 257--265. Springer, 1991.

\bibitem{Courtois:2014uw}
Nicolas~T. Courtois.
\newblock {On The Longest Chain Rule and Programmed Self-Destruction of Crypto
  Currencies}.
\newblock December 2014.
\newblock {Accessed: 2016-07-08}.

\bibitem{nicslide}
Nicolas~T. Courtois.
\newblock {Bitcoin Security: Cryptographic Risks}.
\newblock
  \url{www.nicolascourtois.com/bitcoin/paycoin_catactrypt_51_Paris_03112015.pdf},
  November 2015.
\newblock {Accessed: 2016-08-11}.

\bibitem{courtoisASIC}
Nicolas~T. Courtois.
\newblock {Bitcoin Mining, Internals, Stratum Improvements and Attacks, Forks,
  51\%, Double Spending Attacks}.
\newblock
  \url{http://www.nicolascourtois.com/bitcoin/paycoin_mining_attacks_4.pdf},
  September 2016.
\newblock {Accessed: 2016-10-02}.

\bibitem{Courtois:digsig}
Nicolas~T. Courtois.
\newblock {Digital Signatures and Bitcoin - 3a}.
\newblock \url{http://www.nicolascourtois.com/bitcoin/paycoin_dig_sign_3a.pdf},
  September 2016.
\newblock {Accessed: 2016-10-02}.

\bibitem{rscoin}
George Danezis and Sarah Meiklejohn.
\newblock {Centrally Banked Cryptocurrencies}.
\newblock {\em arXiv preprint arXiv:1505.06895}, 2015.

\bibitem{delerablee2006dynamic}
C{\'e}cile Delerabl{\'e}e and David Pointcheval.
\newblock {Dynamic Fully Anonymous Short Group Signatures}.
\newblock In {\em Progress in Cryptology-VIETCRYPT 2006}, pages 193--210.
  Springer, 2006.

\bibitem{Eyal:2014kd}
Ittay Eyal and Emin~G{\"u}n Sirer.
\newblock {Majority Is Not Enough: Bitcoin Mining Is Vulnerable.}
\newblock {\em Financial Cryptography}, pages 436--454, 2014.

\bibitem{fisher2008finding}
Tom Fisher.
\newblock Finding rational points on elliptic curves using 6-descent and
  12-descent.
\newblock {\em Journal of Algebra}, 320(2):853--884, 2008.

\bibitem{BNCurve}
Pierre-Alain Fouque and Mehdi Tibouchi.
\newblock {Indifferentiable Hashing to Barreto--Naehrig Curves}.
\newblock In {\em LATINCRYPT -- International Conference on Cryptology and
  Information Security in Latin America}, pages 1--17. Springer, 2012.

\bibitem{f2012}
Matthew~K Franklin and Haibin Zhang.
\newblock {A Framework for Unique Ring Signatures.}
\newblock {\em IACR Cryptology ePrint Archive}, 2012:577, 2012.

\bibitem{gallant2001faster}
Robert~P Gallant, Robert~J Lambert, and Scott~A Vanstone.
\newblock Faster point multiplication on elliptic curves with efficient
  endomorphisms.
\newblock In {\em Annual International Cryptology Conference}, pages 190--200.
  Springer, 2001.

\bibitem{gentry2009fully}
Craig Gentry.
\newblock {\em {A Fully Homomorphic Encryption Scheme}}.
\newblock PhD thesis, Stanford University Dissertation, 2009.
\newblock Available at:
  \url{https://crypto.stanford.edu/craig/craig-thesis.pdf}.

\bibitem{Giechaskiel:2016wv}
Ilias Giechaskiel, Cas Cremers, and Kasper~B Rasmussen.
\newblock {On Bitcoin Security in the Presence of Broken Crypto Primitives}.
\newblock February 2016.
\newblock {Accessed: 2016-06-09}.

\bibitem{Gold1985}
Shafi Goldwasser, Silvio Micali, and Charles Rackoff.
\newblock {The Knowledge of Interactive Proof Systems}.
\newblock {\em ACM symposium on theory of computing}, 17:291--304, 1985.

\bibitem{groth2007fully}
Jens Groth.
\newblock Fully anonymous group signatures without random oracles.
\newblock In {\em International Conference on the Theory and Application of
  Cryptology and Information Security}, pages 164--180. Springer, 2007.

\bibitem{Darkcoin}
Kyle Hagan and Evan Duffield.
\newblock {Darkcoin: Peer{\-}to{\-}Peer Crypto{\-}Currency with Anonymous
  Blockchain Transactions and an Improved Proof{\-}of{\-}Work System }.
\newblock In {\em Industrial Engineering and Engineering Management (IEEM),
  2014 IEEE International Conference on}, pages 1443--1447. IEEE, 2014.

\bibitem{hopwood}
Daira Hopwood, Sean Bowe, Taylor Hornby, and Nathan Wilcox.
\newblock {ZCash Protocol Specification, Version 2016.0-alpha-3.1}.
\newblock
  \url{https://github.com/zcash/zips/blob/master/protocol/protocol.pdf}, May
  2016.
\newblock {Accessed: 2016-08-25}.

\bibitem{HashtoEC}
Thomas Icart.
\newblock {How to Hash into Elliptic Curves}.
\newblock In {\em {Advances in Cryptology -- CRYPTO 2009}}, pages 303--316.
  Springer, 2009.

\bibitem{johnson2001elliptic}
Don Johnson, Alfred Menezes, and Scott Vanstone.
\newblock The elliptic curve digital signature algorithm (ecdsa).
\newblock {\em International Journal of Information Security}, 1(1):36--63,
  2001.

\bibitem{tendermint}
Jae Kwon.
\newblock Tendermint: Consensus without mining.
\newblock 2014.

\bibitem{RSK}
Sergio Lerner.
\newblock {RSK Rootstock Platform -- Bitcoin powered smart contracts}.
\newblock Published at \url{http://www.rsk.co/}, November 2015.
\newblock {Accessed: 2016-08-23}.

\bibitem{lindhurst1999analysis}
Scott Lindhurst.
\newblock An analysis of shanks’s algorithm for computing square roots in
  finite fields.
\newblock In {\em Number theory (Ottawa, 1996), CRM Proc. Lecture Notes},
  volume~19, pages 231--242, 1999.

\bibitem{coinjoin}
Gregory Maxwell.
\newblock {CoinJoin: Bitcoin privacy for the real world}.
\newblock \url{https://bitcointalk.org/index.php?topic=279249}, August 2013.
\newblock {Accessed: 2016-08-16}.

\bibitem{CT}
Gregory Maxwell.
\newblock {Confidential Transactions}.
\newblock \url{https://people.xiph.org/~greg/confidential_values.txt}, June
  2015.
\newblock {Accessed: 2016-08-16}.

\bibitem{menezes2012elliptic}
Alfred~J Menezes.
\newblock {\em Elliptic curve public key cryptosystems}.
\newblock KLUWER ACADEMIC PUBLISHERS, 1993.

\bibitem{miers2013zerocoin}
Ian Miers, Christina Garman, Matthew Green, and Aviel~D Rubin.
\newblock Zerocoin: Anonymous distributed e-cash from bitcoin.
\newblock In {\em Security and Privacy (SP), 2013 IEEE Symposium on}, pages
  397--411. IEEE, 2013.

\bibitem{bit}
Satoshi Nakamoto.
\newblock {Bitcoin: A Peer-to-Peer Electronic Cash System}.
\newblock 2008.
\newblock {Accessed: 2016-03-30}.

\bibitem{shnoe}
Shen Noether.
\newblock {Broken Crypto in Shadowcash}.
\newblock
  \url{https://shnoe.wordpress.com/2016/02/11/de-anonymizing-shadowcash-and-oz-coin/},
  February 2016.
\newblock {Accessed: 2016-08-20}.

\bibitem{noether2016ring}
Shen Noether, Adam Mackenzie, and Monero~Core Team.
\newblock {Ring Confidential Transactions}.
\newblock 2016.
\newblock {Accessed: 2016-08-09}.

\bibitem{petit2016algebraic}
Christophe Petit, Michiel Kosters, and Ange Messeng.
\newblock {Algebraic Approaches for the Elliptic Curve Discrete Logarithm
  Problem over Prime Fields}.
\newblock In {\em IACR International Workshop on Public Key Cryptography},
  pages 3--18. Springer, 2016.

\bibitem{poonen2001computing}
Bjorn Poonen.
\newblock Computing rational points on curves.
\newblock {\em Number Theory for the Millenium}, 2001.

\bibitem{2001leak}
Ronald~L Rivest, Adi Shamir, and Yael Tauman.
\newblock {How to Leak a Secret}.
\newblock In {\em {Advances in Cryptology -- ASIACRYPT 2001}}, pages 552--565.
  Springer, 2001.

\bibitem{shallue}
Andrew Shallue and Christiaan~E van~de Woestijne.
\newblock Construction of rational points on elliptic curves over finite
  fields.
\newblock In {\em International Algorithmic Number Theory Symposium}, pages
  510--524. Springer, 2006.

\bibitem{standardnational}
Secure~Hash Standard.
\newblock {National Institute of Standards and Technology (NIST), FIPS
  Publication 180-2, Aug 2002}.
\newblock
  \url{http://csrc.nist.gov/publications/fips/fips180-2/fips180-2withchangenotice.pdf},
  2002.
\newblock {Accessed: 2016-08-16}.

\bibitem{szydlo2004merkle}
Michael Szydlo.
\newblock Merkle tree traversal in log space and time.
\newblock In {\em International Conference on the Theory and Applications of
  Cryptographic Techniques}, pages 541--554. Springer, 2004.

\bibitem{teranishi2004k}
Isamu Teranishi, Jun Furukawa, and Kazue Sako.
\newblock K-times anonymous authentication.
\newblock In {\em International Conference on the Theory and Application of
  Cryptology and Information Security}, pages 308--322. Springer, 2004.

\bibitem{CSNARK}
Eran Tromer.
\newblock {C: There's a SNARK for that}.
\newblock
  \url{https://www.cs.tau.ac.il/~tromer/slides/csnark-usenix13rump.pdf}, 2013.
\newblock {Accessed: 2016-08-21}.

\bibitem{CryptN}
Nicolas van Saberhagen.
\newblock {Cryptonote v 2.0}.
\newblock 2013.
\newblock {Accessed: 2016-08-09}.

\bibitem{yp}
Gavin Wood.
\newblock {Ethereum: A Secure Decentralised Generalised Transaction Ledger --
  Homestead Revision}.
\newblock January 2016.

\bibitem{Zamfir}
Vlad Zamfir.
\newblock {Introducing Casper ``the Friendly Ghost''}.
\newblock Published at
  \url{https://blog.ethereum.org/2015/08/01/introducing-casper-friendly-ghost/},
  August 2015.
\newblock {Accessed: 2016-04-28}.

\bibitem{enig}
Guy Zyskind, Oz~Nathan, and Alex Pentland.
\newblock {Enigma: Decentralized Computation Platform with Guaranteed Privacy}.
\newblock {\em arXiv preprint arXiv:1506.03471}, 2015.

\end{thebibliography}

\newpage \begin{appendices} \section{Homogeneous Elliptic Curve Equations}

Elliptic curves are defined as the coordinate points that satisfy the following
  (Weierstrass) equation.  Formally, for a finite field $ \mathbb{F}_q $, with
  $q$ a prime power, and algebraic closure of $ \mathbb{F}_q $ (which is the
  set $ \bigcup_{m \geq 1} \mathbb{F}_{q^m} $), we have the equation below
  defining an elliptic curve: \begin{equation}\label{WE} Y^2Z + a_1 XYZ + a_3
  YZ^2 = X^3 + a_2 X^2Z a_4 XZ^2 + a_5 Z^3, \end{equation}

with $a_1, a_2, a_3, a_4, a_5 \in \bigcup_{m \geq 1} \mathbb{F}_{q^m}$.  We
  call $\mathbb{F}_q$ the \emph{base field} for the elliptic curve group.

The equation shown in (\ref{WE}) is called the homogeneous or projective
  Weierstrass equation, and is defined up to equivalence in the projective
  plane, $P^2( \mathbb{F}_q )$.  The equivalence is a simple linear relation
  acting on $ ( \mathbb{F}_q ) ^3 \backslash \{ ( 0, 0, 0 ) \} $ with $ ( x_1,
  y_1, z_1 ) \sim ( x_2, y_2, z_2 ) $ if and only if there exists $ u \in
  \mathbb{F}_q^*$ such that $x_1 = ux_2$, $y_1 = uy_2$, and $z_1 = uz_2$.  The
  points $(X, Y, Z)$ satisfying equation (\ref{WE}) above form an equivalence
  class of projective points $(X \colon Y \colon Z)$.

An elliptic curve, formally, is the set of all solutions in $P(\bigcup_{m \geq
  1} \mathbb{F}_q)$ of a \emph{smooth} Weierstrass equation.  To define smooth
  in this context, we first rearrange equation (\ref{WE}), to form:
  \begin{equation}\label{eqtwo} F(X, Y, Z) = Y^2Z + a_1 XYZ + a_3 YZ^2 - X^3 -
  a_2 X^2 Z - a_4 XZ^2 - a_5 Z^3 = 0.  \end{equation}

A smooth curve is one which has no points at which all of $ \frac{\partial
  F}{\partial X}$, $ \frac{\partial F}{\partial Y}$, and $\frac{\partial
  F}{\partial Z}$ vanish.  In other words, at least one partial derivative of
  $F$ must be non-zero at each point satisfying $F(X, Y, Z)$.  The point with
  $Z = 0$, $( 0 \colon 1 \colon 0 )$ acts as the `point at infinity', and in
  \emph{affine} coordinates, is represented by a line at plus and minus
  infinity on the $y$ axis.  This point at infinity acts as the additive zero
  element.

As the curve in homogeneous form is a linear equivalence class, we can divide
  each of $X$, $Y$ and $Z$ by $Z$, thus giving $Z$ as 1, a constant that we
  needn't write down, and with $x = \frac{X}{Z}$ and $y = \frac{Y}{Z}$ we
  produce the affine coordinates of the curve.

  \section{Elliptic Curve Arithmetic \& ECDSA}

Arithmetic in elliptic curve groups does not occur as one may naively assume,
  with $(x_1, y_1) + (x_2, y_2) = (x_1 + x_2, y_1 + y_2)$. Instead, point
  addition, point doubling and scalar point multiplication take place through
  the formulae given below.

\begin{description} \item[Point Addition] For simple point addition, $(x_1,
      y_1) + (x_2, y_2) = (x_3, y_3)$, with $(x_1 \neq x_2)$, we form $(x_3,
      y_3)$ through the following: \begin{align} \text{For } \lambda & =
        \frac{y_2 - y_1}{x_2 - x_1}, \\ x_3 & = \lambda^2 - x_1 - x_2, \\ y_3 &
      = \lambda(x_1 - x_3) - y_1.  \end{align} It is clear here that $\lambda$
      is formed by taking the gradient between the two points, and addition
      formed in accordance with the elliptic curve equation.

\item[Point Doubling] For $y_1 \neq 0$, we set $\lambda = \frac{3x_1 +
  a}{2y_1}$, and use the formulae for $x_3$ and $y_3$ as given above.  We see
    that $\lambda$ here is the tangential gradient of the point we are
    doubling, formed through differentiation of the elliptic curve equation.
    Special cases, such as when $y_1 = 0$, are explained in detail in Menezes'
    Elliptic Curve Public Key Cryptosystems \cite{menezes2012elliptic}.

\item[Scalar Multiplication] Scalar multiplication occurs through repeated
  point doubling, much as exponentiation over the integers $g^x$ can be
    calculated through multiplying $g$ by itself $x$ times.

\item[ECDSA] ECDSA signatures are formed of pairs, $(r, s)$, constructed as
  follows.

With $k$ a random nonce, $r$ is first formed $r = (k \cdot g)_x$ (the $x$
    coordinate of the elliptic curve point given by $k \cdot g$).  $s$ is then
    constructed $s = \frac{z + r \cdot d}{k}$, with $d$ the private key, $z$
    the hash of the signed message authorising the transaction.

If $k$ is known, for example due to low entropy sources of randomness being
    used, we will be able to calculate the private key, as all other
    information is public. Reusing $k$ and $d$ together also gives adversaries
    all information needed to calculate the private key.  This has happened
    several times in the history of bitcoin, leading to losses of up to \$58
    million \cite{nicslide}.  RFC6979 is suggested as a deterministic source of
    randomness, both to increase security against low entropy randomness, and
    against the potential for backdoors to be introduced through manipulated
    sources of randomness \cite{nicslide}.  \end{description}

 \section{The Franklin and Zhang NIZK Proof}

The Franklin and Zhang paper gives the assumption (without loss of generality)
that $log_{H(m||R)} \mathlarger{\mathlarger{\tau}} = log_gy_i$ \cite{f2012}.
This assumption is critical in the proof.  The explanation for its formation is
as follows, starting from construction: \begin{align*} &
  \mathlarger{\mathlarger{\tau}} = H(m||R)^{x_i}, \ y_i = g^{x_i} \\ \implies &
  x_i = log_{H(m||R)} (\mathlarger{\mathlarger{\tau}}) = log_g(y_i) \\ \implies
  & log_{H(m||R)} (\mathlarger{\mathlarger{\tau}}) = log_g(y_i).  \end{align*}
  From this assumption, the following proof system is constructed \cite{f2012}:

\begin{enumerate} \item For $j \in [n]$ and $j \neq i$, prover selects $c_j,
      t_j \leftarrow_R \mathbb{Z}_q$, and computes $a_j \leftarrow
      g^{t_j}y_j^{c_j}$ and $b_j \leftarrow H(m)^{t_j}(H(m)^{x_i})^{c_j}$; for
      $j = i$, prover selects $r_i \leftarrow_R \mathbb{Z}_q$ and computes $a_i
      \leftarrow g^{r_i}$ and $b_i \leftarrow H(m)^{r_i}$. Prover then sends
      the set $ \{a_j, b_j\}^n_1 $ to the verifier.  \item Verifier sends
      challenge, formed $c \leftarrow_R \mathbb{Z}_q$, to the prover.  \item
        Prover computes $c_i \leftarrow c - \sum_{j \neq i} c_j$ and $t
        \leftarrow r - c_i x_i$ mod $q$, sends the pairs $c_1, t_1, \ldots,
        c_n, t_n$ to the verifier.  \item Verifier accepts if and only if $a_j
= g^{t_j}y_j^{c_j}$ and $b_j = H(m)^{t_j} \mathlarger{\tau}^{c_j}$, for all $j
\in [n]$.  \end{enumerate}

  \section{Completeness} We need only prove that a correctly formed signature
  always verifies.  Therefore, our proof seeks to show: $$ \sum^n_1 c_j
  \stackrel{?}{=} H'(m, R, \{g^{t_j}y_j^{c_j}, H(m||R)^{t_j} \tau^{c_j}
  \}^n_1). $$

    By definition, we have $j \neq i$, $c_j \leftarrow_R \mathbb{Z}_q$, $ c_i
    \leftarrow H'(m, R, \{a_j, b_j\}^n_1) - [ \sum_{j \neq i} c_j ] \textrm{
      mod } q$, which gives:

    \begin{equation*} \begin{split} & \sum^n_1 c_j = \sum_{j \neq i}c_j + H'(m,
      R, \{a_j, b_j \}^n_1) - \sum_{j \neq i}c_j, \\ \implies & \sum^n_1 c_j =
      H'(m, R, \{a_j, b_j \}^n_1) = H'(m, R, \{a_j, b_j\}_{j \neq i}, \{a_i,
      b_i\}) \\ \iff & \{ a_j, b_j \}^n_1 = \{a_j, b_j\}_{j \neq i}, \{a_i,
      b_i\} \end{split} \end{equation*}

    For $j \neq i$, by construction we have $a_j = g^{t_j}y_j^{c_j}$, and $b_j
    = H(m||R)^{t_j}(H(m||R)^{x_i})^{c_j}$.

    Using the definition $\tau = H(m||R)^{x_i}$ (constructed with a valid
    secret key $x_i$), we see that the verification equation holds
    unconditionally by definition for $j \neq i$.

    For the case $j = i$, we need to prove the equivalence, $$\{a_i, b_i\} = \{
      g^{r_i}, H(m||R)^{r_i} \} \stackrel{?}{=} \{ g^{t_i}y_i^{c_i},
    H(m||R)^{t_i}\tau^{c_i} \}.$$

    We here must use the construction of $t_i$ in order to show the sides of
    the equation are equivalent. Again this assumes knowledge of a valid secret
    key $x_i$, as $t_i$ is defined $t_i = r_i - c_i x_i$.

    Taking the first element, $a_i$, and using the knowledge that $y_i =
    g^{x_i}$, we see that: $$ g^{r_i} = g^{t_i}g^{x_i c_i} \implies g^{r_i} =
    g^{r_i - c_i x_i}g^{x_i c_i} \implies g^{r_i} = g^{r_i} $$

    The proof of equivalence on the other element works similarly, replacing
    $g$ with $H(m||R)$. $\blacksquare$

    \section{Unforgeability} This proof is completed through a series of games,
    suggested and briefly explored in \cite{f2012}.  We let $W_i$ be the event
    that the adversary, $A$, succeeds in Game $i$.

    \subsection*{Game 0}

    The original unforgeability experiment between challenger and adversary,
    $A$, is described as follows:

    Let $(R^*, m^*, \sigma^*)$ be the output of adversary $A$, let $W_0$ be the
    event that $A$ succeeds in producing a verifiable triple, under the
    conditions that $R^* \subseteq T\textbackslash CU$ (with $T$ defined below
    in the key generation stage), and $CU$ as the set of keys of corrupted
    users, which the adversary has access to.

    In short, this means that the adversary must produce a signature with a key
    pair which they have not been explicitly given.  We also have the condition
    that $A$ has never queried \textbf{RS}$( \cdot, \cdot, \cdot)$ with
    $(\cdot, R^*, m^*)$, and obviously we must have \textbf{Ver}$(R^*, m^*,
    \sigma^*)$ = 1.

    By definition, with $\textbf{Adv}^{uf}_{RS}(A)$ meaning the advantage of
    adversary $A$ in relation to the unforgeability property on the ring
    signing algorithm, and $Pr[w_0]$ as the probability of success in the game
    above, we have: $$ \textbf{Adv}^{uf}_{RS}(A) = Pr[W_0].  $$

    \subparagraph{Generation of $n$ Public Keys} Challenger chooses $x, y
    \leftarrow_R \mathbb{Z}_q $, and computes DDH triple (defined in section 1)
    $(g, X, Y, Z)$, such that $X = g^x$, $Y = g^y$, and $Z = g^{xy}$.  For all
    $i \in [n]$, challenger randomly selects $x_i \leftarrow_R \mathbb{Z}_q$
    and calculates $pk_i \leftarrow X \cdot g^{x_i}$, and gives $T =
    \{pk_i\}^n_{i=1}$ to the adversary $A$.

    We have that: $$\textbf{Adv}^{uf}_{RS}(A) = Pr[W_0].$$

    \subsection*{Game 1}

    \subparagraph{Verification} The final forgery $(R^*, m^*, (\tau^*, \pi^*)
    )$ checks $\exists i \in [n]$ such that $pk_i \in R^*$, and $(g, pk_i$,
    $H(m^*||R^*)^{sk_i}, \tau^*)$ is a DDH triple.  This means that given $g,
    pk_i$ and either $H(m^*||R^*)^{sk_i}$ or $\tau^*$, it is impossible to
    determine which of the latter two elements you have received.

    With $A_1$ an adversary that attacks the adaptive soundness property of the
    underlying NIZK proof system, we have that: $$Pr[W_0] - Pr[W_1] \le
    \textbf{Adv}^{sound}_{(P,V)}(A_1).$$

    This probability is bounded in real terms by $(q_h + 1)/q$, with 1
    occurring only if the adversary did not query the $H'$ oracle during its
    attempted forgery.

    \subsection*{Game 2} Game 2 builds on Game 1 to include a simulator $S$ to
    simulate the NIZK proof of queried signatures.  We have here that, for an
    adversary that attacks the \emph{adaptive security} of the NIZK proof in
    Franklin and Zhang's `Framework for Unique Ring Signatures': $$ Pr[W_1] -
    Pr[W_2] \le \textbf{Adv}^{zk}_{(P,V)}(A_2)$$ \subsection*{Game 3} Game 3 is
    an adaptation of Game 2, with the DDH triple provided replaced with a
    random tuple. Any adversary that can distinguish between the DDH and random
    triple can be converted into an adversary, here called $A_3$, who can solve
    the DDH problem. That is, $$ Pr[W_2] - Pr[W_3] \le
    \textbf{Adv}^{ddh}_{\mathbb{G}}(A_3).$$ Due to this, we have here that
    $Pr[W_3] \le n/q$.

    The following result holds, concluding the unforgeability proof: $$
    \textbf{Adv}^{uf}_{RS}(A) \le \textbf{Adv}^{ddh}_{\mathbb{G}}(A_3) + (2q_h
    + n + 1)/q.$$

    \section{Anonymity} This work is an adaptation of the proofs presented in
    `A Framework for Unique Ring Signatures' \cite{f2012}.  The proof takes
    place through a series of three games, defined below.  \subsection*{Game 0}
    Let Game 0 be the original anonymity experiment between challenger and
    adversary, $A$.  Assume $A$ makes at most $q_h$ queries to either $H$ or
    $H'$, and at most $q_s$ queries to the signing oracle as defined below.
    The game takes place in a setting where the adversary is given the ability
    to query the hash and signing oracles, and must distinguish between
    correctly formed and random chosen triples.

    \subparagraph{Generation of $n$ Public Keys} Challenger chooses $x, y
    \leftarrow_R \mathbb{Z}_q $, and computes DDH triple (defined in section 1)
    $(g, X, Y, Z)$, such that $X = g^x$, $Y = g^y$, and $Z = g^{xy}$.  For all
    $i \in [n]$, challenger randomly selects $x_i \leftarrow_R \mathbb{Z}_q$
    and calculates $pk_i \leftarrow X \cdot g^{x_i}$, and gives $T =
    \{pk_i\}^n_{i=1}$ to the adversary $A$.  We see all public keys here have
    been chosen independently, at random.

    \subparagraph{Queries to $H$} Challenger maintains set $V$, constructed
    $(m, R, h, u)$, initially empty.  Responses to hash queries (of the form
    $(m_j, R_j)$) are not as straight forward as simply hashing the input -
    instead, the hash queries are formed like queries to a random oracle
    constructed in the following way: \begin{itemize} \item Challenger randomly
        selects $d \leftarrow_R \mathbb{Z}_q$.  \item In response to a hash
          query on $(m_j, R_j)$, challenger checks if $ \exists (m_j, R_j, h_j,
          u_j) \in V$ for any $h_j, u_j$.  \item If so, challenger returns
          $h_j$.  \item If not, challenger randomly selects $u_j \leftarrow_R
            \mathbb{Z}_q$, constructs $h_j \leftarrow Y^d \cdot g^{u_j}$,
            returns $h_j$ and adds $(m_j, R_j, h_j, u_j)$ to set $V$.
    \end{itemize}

    \subparagraph{Queries to $H'$} Challenger maintains set $V'$, constructed
  $(m, R, \{a_j, b_j\}^n_1, c)$, initially empty.  Responses to hash queries
(of the form $(m', R', \{a'_j, b'_j\}^n_1)$) work as follows: \begin{itemize}
  \item Challenger checks if there exists $(m', R', \{a'_j, b'_j\}^n_1, c')$ in
    $V'$.  \item If so, challenger returns $c'$.  \item If not, challenger
picks random $c' \leftarrow_R \mathbb{Z}_q$, returns this $c'$, and adds  $(m',
R', \{a'_j, b'_j\}^n_1, c')$ to $V'$.  \end{itemize}

    \subparagraph{Signing Queries} Signing queries consider two encapsulated
  adversarial parties, as follows.  \begin{itemize} \item Adversary $A$ queries
      the signing oracle with an input of the form $(j, R, m)$.  \item
        Adversary $B$ queries $H$ and receives $h$ constructed $h \leftarrow
        Y^d \cdot g^u$.  \item Challenger computes $\tau \leftarrow Z^d \cdot
        X^u \cdot Y^{dx_j} \cdot g^{x_ju}$, computes corresponding NIZK proof
      $\pi$, using secret key of user $j$.  \item Challenger returns $(m, R,
        \tau, \pi)$ to $A$.  \end{itemize}

    \subparagraph{Challenge} Adversary $A$ requests challenge $(i_0, i_1, R^*,
    m^*)$, with $m^*$ to be signed with respect to ring $R^*$, and $i_0, i_1
    \in [n]$ indices such that $pk_{i_0}, pk_{i_1} \in T \cap R^*$ Define T in
    the Anonymity games so that it's known already here. Or define here.
    Challenger randomly chooses bit $b \leftarrow_R \{0,1\}$, returns challenge
    signature \textbf{RS}$(sk_{i_b}, R^*, m^*)$ to $A$. $A$ cannot query
    \textbf{RS}$(\cdot, \cdot, \cdot)$ with $(i_0, R^*, m^*)$ or $(i_1, R^*,
    m^*)$.

    \subparagraph{Output} Adversary $A$ outputs $b'$ as its guess.

    \subsection*{Game 1} Game 1 is similar to Game 0, but when responding to a
    signing query (j, m, R), the challenger simply simulates the proof.  $c_1,
    t_1, \ldots, c_n, t_n$ are chosen randomly from $\mathbb{Z}_q$, and $a_j =
    g^{t_j}y_j^{c_j}$ and $b_j = H(m||R)^{t_j} \tau^{c_j}$ are computed for
    each $j \in [n]$.  We let $W_1$ be the event of $A$'s success at Game 1.
    It is obvious to deduce that if there exists an adversary $A_1$ that can
    attack the adaptive NIZK property of the fundamental NIZK proof system
    $(P,V)$, the following holds: $$ Pr[W_0] - Pr[W_1] \le
    \textbf{Adv}^{zk}_{(P,V)}(A_1).  $$ \subsection*{Game 2} In Game 2, the
    signing oracle is modified, such that the DDH triple is replaced by a
    random triple.  Adversary $A$ can only notice the difference with
    negligible probability, under the DDH assumption.  Specifically, the
    challenger simply replaces $Z$ with some $\hat{Z}$ where $\hat{Z} = g^c$
    and $c \leftarrow_R \mathbb{Z}_q$.  Let $W_2$ be the event that $A$
    succeeds in Game 2.  We can show that there exists an adversary $A_2$ such
    that: $$Pr[W_1] - Pr[W_2] \le \textbf{Adv}^{ddh}_{\mathbb{G}}(A_2), \ \ \ \
    \ Pr[W_2] = 0.5.$$

    Finally, we can combine the equations produced for each individual game, to
    give the result: $$ \textbf{Adv}^{anon}_{RS}(A) \le
    \textbf{Adv}^{ddh}_{\mathbb{G}}(A_2) + q_h/q.$$

 \end{appendices}

\end{document}